\documentclass[journal=jctc,manuscript=article,layout=twocolumn]{achemso}

\usepackage{chemformula} 
\usepackage[T1]{fontenc} 
\usepackage{textgreek}
\usepackage{siunitx}
\usepackage{soul}
\usepackage{amssymb}
\usepackage[utf8]{inputenc}

\author{Zakarya Benayad}
\affiliation{CPCV, Département de Chimie, École Normale Supérieure, PSL University, Sorbonne University, CNRS, 75005 Paris, France}
\author{Guillaume Stirnemann}
\affiliation{CPCV, Département de Chimie, École Normale Supérieure, PSL University, Sorbonne University, CNRS, 75005 Paris, France}
\email{guillaume.stirnemann@ens.psl.eu}

\title{Hamiltonian replica exchange augmented with diffusion-based generative models and importance sampling to assess biomolecular conformational basins and barriers}

\begin{document}
		
\begin{abstract}
Enhanced sampling techniques are essential for exploring biomolecular conformational dynamics that occur on timescales inaccessible to conventional molecular dynamics (MD) simulations. This study introduces a framework that combines Hamiltonian replica exchange with solute tempering (REST2) with denoising diffusion probabilistic models (DDPMs) and importance sampling to enhance the mapping of conformational free-energy landscapes. Building on previous applications of DDPMs to temperature replica exchange (TREM), we propose two key improvements. First, we adapt the method to REST2 by treating potential energy as a fluctuating variable. This adaptation allows for more efficient sampling in large biomolecular systems. Second, to further improve resolution in high-barrier regions, we develop an iterative scheme combining replica exchange, DDPM, and importance sampling along known collective variables. Benchmarking on the mini-protein CLN025 demonstrates that DDPM-refined REST2 achieves comparable accuracy to TREM while requiring fewer replicas. Application to the enzyme PTP1B reveals a loop transition pathway consistent with prior complex biased simulations, showcasing the approach's ability to uncover high-barrier transitions with minimal computational overhead with respect to conventional replica exchange approaches. Overall, this hybrid strategy enables more efficient exploration of free-energy landscapes, expanding the utility of generative models in enhanced sampling simulations.
\end{abstract}

\section{Introduction}

Despite decades of advancements in software and hardware, molecular dynamics (MD) simulations\cite{tuckerman2023statistical} of biomolecules remain significantly constrained by accessible timescales, which are often much shorter than those relevant to experiments. While simulations now routinely reach the microsecond timescale, processes occurring on much longer timescales remain inaccessible to brute-force MD. To study such events, one must rely on enhanced sampling methods\cite{henin2022enhanced}, which accelerate transitions along the system's slow degrees of freedom.

Enhanced sampling techniques generally fall into three categories, each with distinct advantages and limitations. Biased sampling methods require identifying a subset of collective variables (CVs) that capture the system’s slow modes. Sampling is then biased using techniques such as umbrella sampling\cite{torrie1977nonphysical}, metadynamics and its variants\cite{laio2002escaping,barducci2008well,invernizzi2020rethinking,bussi2020using}, or selective acceleration techniques (adiabatic free-energy dynamics\cite{10.1063/1.1448491}, temperature-accelerated MD\cite{MARAGLIANO2006168}, etc.), to name a few representative examples. If the chosen CVs accurately describe the reaction coordinates and other degrees of freedom relax quickly relative to the simulation timescale, these methods can, in principle, allow the study of any process, regardless of its free-energy barrier.

Path-ensemble sampling techniques  focus specifically on transition regions\cite{bolhuis2002transition,dellago1998efficient,CHONG201788}, where standard MD simulations spend only a small fraction of the time. While effective in capturing rare events, these methods often require a reasonable initial guess for the transition pathway (and thus even a vague idea about a relevant set of reaction coordinates), making initialization nontrivial. Furthermore, reconstructing the underlying kinetics and thermodynamics is challenging, and the generated transition path ensembles tend to be highly correlated, reducing sampling efficiency. We note however that continuous improvement of these methods has resulted in strategies that are more robust against the above pitfalls, such as replica-exchange transition interface sampling \cite{PhysRevLett.98.268301,doi:10.1073/pnas.2318731121}. 

Generalized ensemble sampling provides a third alternative by simulating the system within an altered ensemble where slow transitions are artificially accelerated. This acceleration can be achieved by modifying e.g. the system’s temperature\cite{sugita1999replica} or potential energy\cite{wang2011replica}. Expanded ensemble techniques such as simulated tempering\cite{marinari1992simulated} and Gaussian-accelerated MD\cite{10.1063/1.1755656} operate with a single system copy, but thermodynamic properties of the unperturbed system are often obtained using replica-based strategies. In these schemes, multiple replicates of the system experience a ladder of perturbations and periodically exchange configurations, improving sampling and facilitating the reconstruction of thermodynamic properties\cite{sugita1999replica,rev2024,wang2011replica}.

These methods are, in principle, agnostic to the slow degrees of freedom of a system, which is a significant advantage over biased sampling and even path-ensemble techniques that typically require some prior knowledge of the reaction pathway\cite{henin2022enhanced}. However, this advantage also comes with drawbacks. Because the biasing effort is distributed across many modes of the system, the sampling of each mode is less efficient than in biased sampling, where a single, well-chosen collective variable (CV) is targeted. As a result, large energy barriers cannot be as easily overcome, as we recently illustrated on realistic and complex biomolecular systems\cite{LANGUINCATTOEN20232744,10.1093/nar/gkz1071}, making these methods less effective for rare-event sampling when compared to direct biasing along a well-defined CV.

Another challenge is that while the exploration of metastable states is improved, the accurate determination of free-energy barriers remains difficult. Even rescaled or high-temperature replicas often suffer from poor statistics in undersampled, high free-energy regions of phase space, and convergence is not easily assessed\cite{henin2022enhanced}. Additionally, these simulations require dozens of replicas, if not more, which represents a significant computational cost.

One way to leverage the information contained in higher-temperature replicas, beyond simply providing configurations to the unscaled reference replica, is through reweighting techniques such as T-WHAM\cite{gallicchio2005temperature} and MBAR\cite{shirts2008statistically}. More recently, Wang et al. introduced a generative approach based on Denoising Diffusion Probabilistic Models (DDPMs) to refine Temperature Replica Exchange Method (TREM) simulations\cite{wang2022data}. DDPM is a stochastic flow-based method that applies discrete forward and backward diffusion steps between the complex probability distribution in the system’s physical space and a more tractable prior distribution\cite{ho2020denoising} in the same space. By learning these noising/denoising processes from imperfect distributions sampled at multiple temperatures, the model can generate new configurations with accurate Boltzmann weights and map them back to the system’s probability distribition\cite{wang2022data}. This approach illustrates a growing trend in the field blending deep-learning methods (including flow-based and diffusion-based generative models) and enhanced sampling molecular simulations\cite{annurev:/content/journals/10.1146/annurev-physchem-083122-125941}. 
\begin{figure*}[t!]
\centering
\includegraphics[width=\textwidth]{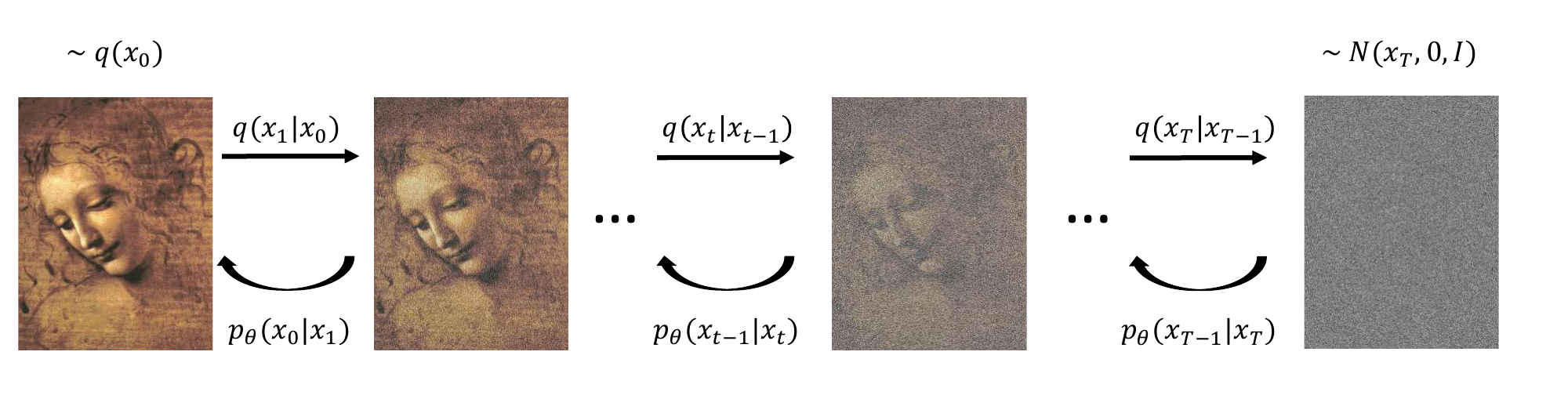} 
\caption{Principles of DDPM. From left to right, the forward diffusion process progressively adds noise, transforming the original data into Gaussian white noise. From right to left, deep neural networks approximate the reverse diffusion process, which maps Gaussian white noise back into meaningful data.}
\label{fig:DDPM_principles}
\end{figure*}

Wang et al. demonstrated that DDPM can learn the joint probability distribution in temperature and configuration space from TREM simulations, enabling the generation of new data points with improved accuracy\cite{wang2022data}. Notably, DDPM outperformed MBAR\cite{shirts2008statistically}, particularly in computing free-energy profiles at temperatures beyond the simulated range. More recent score-based approaches\cite{song2021scorebasedgenerativemodelingstochastic}, which refine the diffusion process in a continuous framework, further highlight the potential of generative models for exploring free-energy landscapes\cite{doi:10.1073/pnas.2321971121}.  A very recent study has also demonstrated their utility (albeit not in a TREM approach) in understanding the conformational free-energy landscapes of intrinsically disordered proteins\cite{Bera2025.01.16.633470}.

However, the potential of generative models should not be overstated. These methods do not inherently solve the sampling problem and remain highly sensitive to the probability distribution(s) on which they are trained. If a metastable state is not at least partially explored in the simulation data, the generative model will not be able to discover or sample it\cite{annurev:/content/journals/10.1146/annurev-physchem-083122-125941}. Consequently, their effectiveness relies heavily on the quality of  sampling in the replica exchange simulations.

In that sense, the current key limitations of the DDPM-TREM approach are twofold: first, TREM becomes impractical and computationally expensive for large systems\cite{stirnemann2015recovering,https://doi.org/10.1002/prot.22702}, and second, large free-energy barriers remain undersampled without additional efforts\cite{LANGUINCATTOEN20232744,10.1093/nar/gkz1071}. To address these challenges, we propose two key improvements. First, we extend the DDPM framework to the Hamiltonian replica exchange scheme REST2\cite{wang2011replica}, which scales better with system size and requires less replicas, and that is regularly used by our group  to enhance conformational exploration in biomolecular systems\cite{https://doi.org/10.1002/chem.202003018,dabin_toward_2023,dabin_atomistic_2024,forget2024simulation}. We demonstrate that DDPM can learn the joint probability distribution in configuration and rescaled potential energy, performing as effectively as TREM in resolving the conformational free-energy surface. Second, we tackle the sensitivity of DDPM to training data quality, particularly in high-barrier regions corresponding to rare events. For systems where relevant reaction coordinates can be identified, we introduce an iterative scheme that combines replica exchange simulation, DDPM, and importance sampling to progressively refine free-energy surfaces and, in particular, free-energy barriers. We demonstrate the effectiveness of our method with two systems of increasing complexity: the chignolin mini-protein CLN025 as a benchmark test case, and the phosphatase PTP1B. In this last case, we show that our approach can uncover a complex transition pathway between two conformations of an active site loop that has only be characterized very recently using extensive biased approaches\cite{zinovjev2024activation}. 

\section{Methods}

\subsection{Basic concepts of DDPM}

Denoising Diffusion Probabilistic Models \cite{ho2020denoising,sohl2015deep} use a stochastic mapping from the empirical complex distribution to a simple prior distribution. DDPMs are inspired by non-equilibrium thermodynamics and are based on two diffusion processes (Figure \ref{fig:DDPM_principles}): a forward diffusion process, also called noising, consisting of a fixed Markov chain that progressively adds noise to the data, and a reverse diffusion process, also called denoising, that learns the inverse mapping from noise to the structured data. \par

\textbf{Forward diffusion process.}
Let $\mathbf{x_{0}}$ denote a sample from the original data (consisting of Cartesian coordinates or a set of given collective variables), drawn from the complex probability distribution $q(\mathbf{x_{0}})$. The forward diffusion process progressively adds noise in $T$ diffusion steps, which constitutes a Markov chain made of $T$ states, $\mathbf{x_{t}}, t\in\{0,..,T\}$, that becomes more and more noisy as $t$ increases. The amount of noise added at each step is fixed by a variance schedule $\beta_{1},  ..., \beta_{T}$. The transition probabilities are Gaussian and are defined as follows:
\begin{equation}
    q(\mathbf{x_{t}}|\mathbf{x_{t-1}}) = \mathcal{N}\left(\mathbf{x_{t}} ; \sqrt{1 - \beta_{t}}\mathbf{x_{t-1}} , \beta_{t}\mathbf{I}\right)
\end{equation}
Hence, as we proceed with the diffusion process, more noise is added to the data, until $x_{T}$, which follows an isotropic Gaussian distribution $\mathcal{N}\left(\mathbf{x_{T}} ; \mathbf{0} , \mathbf{I}\right)$.  \par
\textbf{Reverse diffusion process.} The goal now is to recover a true sample, following the original probability distribution $q(\mathbf{x_{0}})$, from a Gaussian noise sample $x_{T}$, i.e., learn the reverse transition probabilities $p(\mathbf{x_{t-1}}|\mathbf{x_{t}})$. For this, we approximate the real (unknown) transition probabilities $p(\mathbf{x_{t-1}}|\mathbf{x_{t}})$ by a model $p_\theta(\mathbf{x_{t-1}}|\mathbf{x_{t}})$, which is represented by a deep neural network. The reverse diffusion process, mapping a Gaussian noise sample to a true data sample is then given by:
\begin{equation}
    p_\theta(\mathbf{x_{t-1}}|\mathbf{x_{t}}) = \mathcal{N}\left(\mathbf{x_{t-1}} ; \mathbf{\mu_{\theta}}(\mathbf{x_{t}} , t), \mathbf{\sigma_{\theta}}(\mathbf{x_{t}} , t)\right)
\end{equation}
where $\mathbf{\mu_{\theta}}(\cdot, t)$ (mean) and $\mathbf{\sigma_{\theta}}(\cdot , t)$ (standard deviation) are learnable parameters.  The deep neural networks used to represent the approximated reverse transition probabilities, $p_\theta(\mathbf{x_{t-1}}|\mathbf{x_{t}})$ are trained by minimizing the difference between the true reverse process and the prediction of the model. This is done by minimizing the following loss function:

\begin{multline}
    L = \mathbb{E}_{q} [ D_{KL}(q(x_{T}|x_{0})||p(x_{T})) +  \\
   \sum_{t>1}D_{KL}(q(x_{t-1}|x_{t}, x_{0}||p_{\theta}(x_{t-1}|x_{t})) \\
    \ln{p_{\theta}(x_{0}|x_{1})} ]\\
\end{multline}
where $D_{KL}$ is the Kullback-Leibler divergence, which quantifies the difference between probability distributions.

As discussed in ref. \cite{wang2022data}, the generation process can be conditioned to the temperature of interest $T$.. In practice, we propose a slight change in which we generate points in the CV-space, conditioning them in the temperature-space to reproduce the distribution of "atomistic" temperatures (obtained from the atomistic kinetic energies) seen at $T$ in each replica (which is typically Gaussian-distributed with a variance that scales as $T$). A similar conditioning on the generation process is applied when dealing with the potential energy instead of the kinetic energy in the REST2 framework (see below).

\subsection{Architecture and hyperparameters}

We used the same design as presented in \citet{wang2022data}, based on \citet{ho2020denoising}. The deep neural network has a U-Net structure, which is a convolutional neural network consisting of two parts: an encoder (down-sampling path) and a decoder (up-sampling path), with skip connections. We used four down-sampling and four up-sampling residue blocks. The diffusion process consisted of 1000 steps. The network parameters were optimized using the Adam optimizer with a learning rate of $2 \times 10^{-5}$, and an exponential moving average with a decay rate of 0.995. 

\subsection{DDPM error bars}

There are several possible ways to define the uncertainties of the DDPM results. First, we trained multiple models on the same datasets and computed the variability among their predictions. In all cases, the resulting errors were very small, at most 0.1~kcal/mol, demonstrating that the intrinsic uncertainties of the model are negligible. The reported error bars instead correspond to the standard deviations obtained from models trained on successive dataset blocks, analogous to the procedure used for estimating raw simulation errors. These deviations are typically comparable to those observed in the underlying data, and therefore reflect the sampling uncertainties of the training data rather than model error per se.

\subsection{Hamiltonian replica exchange with solute tempering REST2}

TREM becomes impractical for bigger systems, because the probability of coordinate exchanges between neighboring replicas scales as the square root of the number of degrees of freedom. One way to address this issue is by rescaling the potential energy component of the Hamiltonian, possibly on a fraction of the system (which is of course impossible when working with temperature exchange). A popular framework is that of solute tempering, of which the version of \citet{wang2011replica}, called REST2 (Replica Exchange with Solute Tempering 2), is an efficient implementation. In REST2, all replicas are run at the same temperature but on different rescaled potential energy surfaces, rescaling only the solute-solute and solute-water interactions, while keeping the water-water interactions unchanged.  \par
The potential energy function for replica $i$ can be written as: 
\begin{equation}
    E_{i}(\mathbf{X}) = \lambda_{i}E_{pp}(\mathbf{X}) + \sqrt{\lambda_{i}}E_{pw}(\mathbf{X}) + E_{ww}(\mathbf{X})
\end{equation}
where $\mathbf{X}$ is the atomic configuration, $\lambda_{i}$ is a fixed parameter that measures the strength of rescaling in replica $i$, $E_{pp}$ is the solute-solute interaction, $E_{pw}$ is the solute-water interaction, and $E_{ww}$ is the water-water interaction (the solute is typically a protein, hence the $p$ subscript). By analogy with TREM replica exchange, it is often convenient to talk about a system effective temperature $\beta_i = \lambda_i \times \beta_{ref}$ where $\beta_{ref}$ is the reference physical temperature of the system. Since the probability of exchanges scales as the square root of the number of degrees of freedom of only the solute (and not the total system), REST2 scales better with system size compared to TREM. 

\subsection{Free-energy differences between conformational basins}

To compute free-energy differences between basins, we defined boundaries for each basin on the 2D FES and summed the normalized histogram probabilities within each region to the relative probabilities in each state, from which we could finally obtain the free-energy differences. 

\subsection{Simulation strategy (chignolin)}

\subsubsection*{MD set-up} The system was built and simulated with GROMACS 2019.4 \cite{gromacs20194}.  We used the NMR structure obtained from the Protein Data Bank (2RVD) \cite{honda2008crystal} of the mutant chignolin CLN025 (YYDPETGTWY). The Amber99sb-ILDN force field \cite{lindorff2010improved} was used, and the system was solvated in a cubic box, containing 1947 TIP3P water molecules and two sodium ions. The minimization was done using the steepest-descent algorithm. A short NVT equilibration (T = 300 K) was performed for 100 ps, followed by a 2-ns NPT equilibration (T = 300 K, P = 1 bar) with a Parrinello-Rahman barostat \cite{parrinello1981polymorphic,nose1983constant}. The final structure was used as the starting structure for all the replicas of the TREM and REST2 simulations. Simulations were performed in the NVT ensemble using the velocity rescaling  thermostat with a stochastic term \cite{bussi2007canonical}. The time step was set to 2 fs. We followed the procedure described in \cite{stirnemann2015recovering}: for REST2 simulations, we ran 12 replicas at the reference temperature T = 300 K, with $\lambda$ values yielding temperatures ranging from 290 K to 529 K; for TREM, we ran 32 replicas, at temperatures ranging from 290 K to 600 K. Exchanges were attempted every 5000 timesteps, i.e., every 10 ps. Useful examples and guidelines for running REST2 and TREM simulations with GROMACS can be found in the PLUMED tutorial ecosystem \cite{Tribello:2025aa}.

\subsubsection*{Collective variables}
We used three collective variables: the RMSD of the backbone with respect to the folded structure (taken as the final structure of the NPT equilibration), the radius of gyration computed on the C$_{\alpha}$, and the fraction of native contacts $Q$, computed following \cite{lindorff2011fast}. We used these three collective variables to train a DDPM.

\subsubsection*{Umbrella sampling simulations}

We performed Umbrella Sampling simulations along the RMSD, defining 19 windows to ensure good overlap between them. The stiffness constant was set to either 20000 kJ/mol/nm$^2$ or 40000 kJ/mol/nm$^2$ depending on the window, and simulations were run for 300 ns for each window. The last 150 ns were considered for analysis. Constraints were applied using Plumed version 2.7.4 \cite{plumed2019promoting,tribello2014plumed} patched with GROMACS. The free-energy profile along the RMSD was computed using weighted histogram analysis method \cite{kumar1992weighted} as implemented in the Grossfield code \cite{grossfield2023wham}.  

\subsubsection*{DDPM training}
We trained the DDPM using the joint probability distribution of RMSD, $R_{g}$, $Q$ and temperature for TREM, and rescaled potential energy for REST2. We used the last 400 ns, leaving 200 ns for equilibration. This amounted to a total 4000 $\times$ 32 = 128 000 training points for TREM and 4000 $\times$ 24 = 96 000 training points for REST2. We then generated a total of 6 000 000 points per replica, that we used to compute the free-energy profiles in Figure \ref{fig:CLN_comp_energies}. 

\subsection{Simulation strategy (PTP1B)}
\subsubsection*{MD set-up}

The system was built from the closed conformation provided by the authors of \cite{zinovjev2024activation} on their Zenodo repository. We used the Amber99sb-ILDNforce field \cite{lindorff2010improved}. The system was solvated with 15298 TIP3P water molecules, 52 sodium ions, and 46 chloride ions, and the energy was minimized using the steepest-descent algorithm. A short NVT equilibration at T = 300 K was performed for 100 ps, followed by a 2-ns NPT equilibration at T = 300 K and P = 1 bar with a Parrinello-Rahman barostat \cite{parrinello1981polymorphic,nose1983constant}. The final structure was used to start all the replicas of the REST2 simulations. The REST2 simulations were performed in the NVT ensemble using the velocity rescaling  thermostat with a stochastic term \cite{bussi2007canonical}, at a reference temperature T = 300 K and in a cubic box of 8 nm. We used 24 replicas with temperature ranging from 290 K to 600 K. Initial REST2 simulations were run for 600 ns with Gromacs 2022.4 \cite{gromacs20224} patched with Plumed version 2.8.1 \cite{plumed2019promoting,tribello2014plumed}. Exchanges were attempted every 5000 timesteps, ie every 10 ps. 

\subsubsection*{Collective variables}
We used four collective variables to describe the transition from closed to open conformation, following ref. \cite{zinovjev2024activation}: two torsional angles $\psi_{181}$, $\phi_{182}$, and two distances describing the formation and rupture of salt bridges Asp181C$\gamma$-Arg221C$\zeta$  ($d_{181-221}$), and Asp181C$\gamma$-Arg112C$\zeta$  ($d_{181-112}$). 

\subsubsection*{DDPM training}

We trained the DDPM using the joint probability distribution of the four collective variables defined above $\psi_{181}$, $\phi_{182}$, $d_{181-112}$, $d_{181-221}$, and the rescaled potential energy function. For training the DDPM on the unbiased REST2 simulations, we used the last 400 ns out of 600 ns, and for the DDPM on the biased REST2 simulations, we used the last 250 ns out of 300 ns. We generated 6,000,000 structures per replica.

\section*{Results and Discussion}

\subsection{DDPM-REST2 implementation}

We first demonstrate how we can extend the approach of \citet{wang2022data} to deal with data from REST2 simulations. In REST2, all replicas are run at the same reference temperature, but on different potential energy surfaces. The potential energy of a configuration $\mathbf{X}$ in replica $i$, is defined as :
\begin{equation}
    E_{i}(\mathbf{X}) = \lambda_{i}E_{pp}(\mathbf{X}) + \sqrt{\lambda_{i}}E_{pw}(\mathbf{X}) + E_{ww}(\mathbf{X}). 
\label{eq:REST2_pot_nrj}
\end{equation}
In their formulation of DDPM applied to TREM\cite{wang2022data}, Wang et al. proposed to consider temperature as a random fluctuating variable used to learn (and improve) a joint probability distribution in physical space and temperature. This was trivially achieved using  the instantaneous temperature computed from the kinetic energy \cite{wang2022data}. 
For REST2, we propose to treat the potential energy as a random fluctuating variable, and to learn the corresponding joint probability distribution in space and potential energy. In practice, we tested two variants in which we considered either the total potential energy $E_{i}$, or, the rescaled part of it $E_{rescaled,i}(\mathbf{X})=\lambda_{i}E_{pp}(\mathbf{X}) + \sqrt{\lambda_{i}}E_{pw}(\mathbf{X})$. 

\begin{figure}[t!]
\centering
\includegraphics[width=0.7\linewidth]{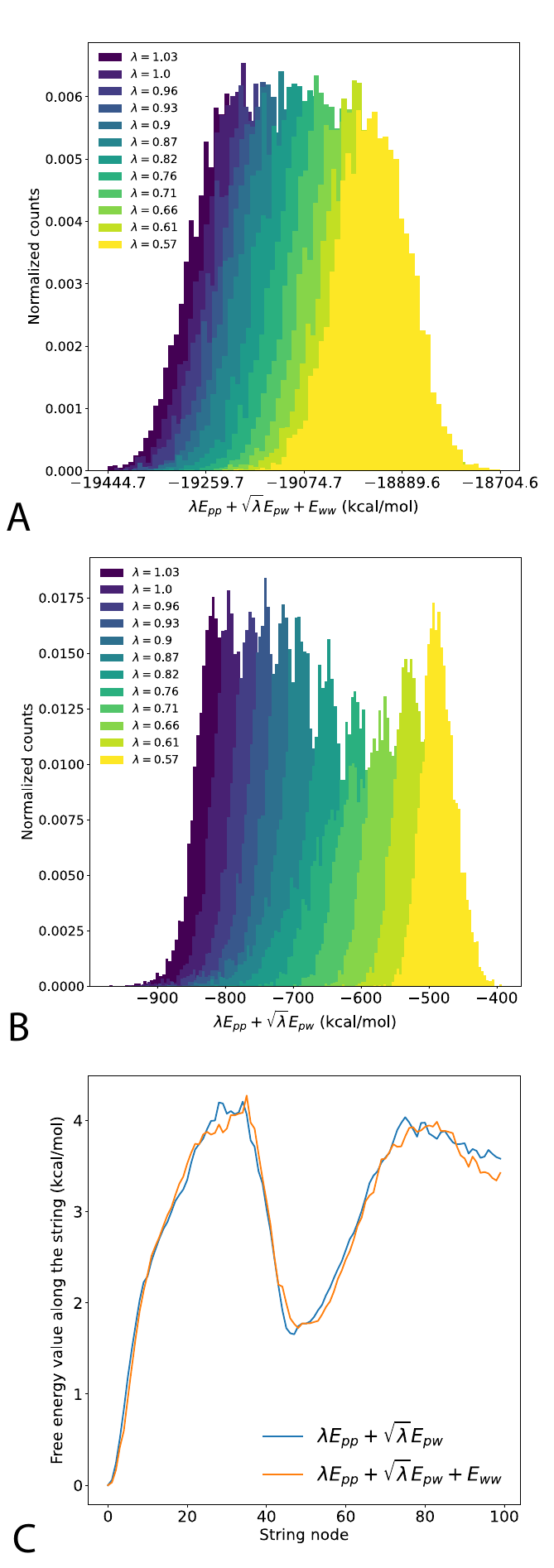} 
\caption{Comparison of different energy terms considered for DDPM. (A) Distributions of the total potential energy for the different replicas. (B) Distribution of $E_{rescaled,i}(\mathbf{X})=\lambda_{i}E_{pp}(\mathbf{X}) + \sqrt{\lambda_{i}}E_{pw}(\mathbf{X})$ for the different replicas. (C) Value of the free-energy surface taken along the minimum free-energy path from the prediction of DDPM trained on the total potential energy and on $E_{rescaled,i}$.}
\label{fig:CLN_comp_energies}
\end{figure}

To benchmark the method, we first present results on the mini-protein chignolin CLN025. REST2 simulations using 12 replicas of 600 ns ranging from 290 K to 530 K ($\lambda_i$ values ranging from 1.03 to 0.57) were performed at a reference physical temperature of 300 K, yielding an exchange rate of 0.51. 

Although DDPM can handle dozens of collective variables\cite{wang2022data,doi:10.1073/pnas.2321971121}, for large systems some choice need to be made as these cannot encompass e.g. all backbone degrees of freedom. Therefore, even for this small protein for which many more CVs could be used, we trained the DDPM on three CVs deemed important for folding: the RMSD with respect to the folded structure (computed on the backbone atoms), the radius of gyration $R_{g}$ (computed on the $C_{\alpha}$), and the fraction of native contacts $Q$\cite{lindorff2011fast}. As a consequence, what DDPM generates are points (and thus probability distributions) in this reduced CV-space, and not 3D Cartesian coordinates of the full biomolecule.

To compute $E_{rescaled,i}$, we need to estimate $E_{pp}$, $E_{pw}$ and $E_{ww}$. For this, we use the \textit{rerun} command of Gromacs to recalculate the energies along the trajectory of each replica with the different Hamiltonians, which gives for a given structure its potential energy as a function of $\lambda$. We then fit $E_{\lambda_i}(\mathbf{X})$ as a function of $\lambda_i$ to obtain the three unknown variables $E_{pp}$, $E_{pw}$ and $E_{ww}$ for this configuration using Equation~\ref{eq:REST2_pot_nrj}. Note that other approaches to compute $E_{pp}$, $E_{pw}$ and $E_{ww}$ directly (for example by calculating the energies for groups of atoms containing either only the protein or the solvent) could suffer from issues related to the typical net charge of the protein because of the Ewald summation on a periodic, charged system.

We find that $E_{rescaled,i}$ is Gaussian distributed within a given trajectory, analogous to the kinetic energy (and hence the instantaneous temperature) used in \citet{wang2022data}, and like the total potential energy function (Figure \ref{fig:CLN_comp_energies}A and B). By computing the minimum free-energy path on the two-dimensional free-energy surface computed along RMSD and $R_{g}$ (see Figure~\ref{fig:CLN_comp_REST2_TREM} and discussion below), we compare the predictions of DDPM trained with the rescaled part of the potential energy, and the total potential energy (Figure \ref{fig:CLN_comp_energies}C), finding no significant difference (see Methods for details about the training and data generation procedure). We decided to use the former for the rest of this work. 


\subsection{Comparison with TREM and US}

We now provide a critical comparison between the results of DDPM-REST2 and that of DDPM-TREM using a similar simulation setup. We performed TREM simulations using 32 replicas of 600 ns ranging from 279 K to 600 K, yielding an exchange rate of 0.13 (this illustrates the much lower performance of TREM compared to REST2 despite three times more replicas). 

\begin{figure}[t!]
\centering
\includegraphics[width=\linewidth]{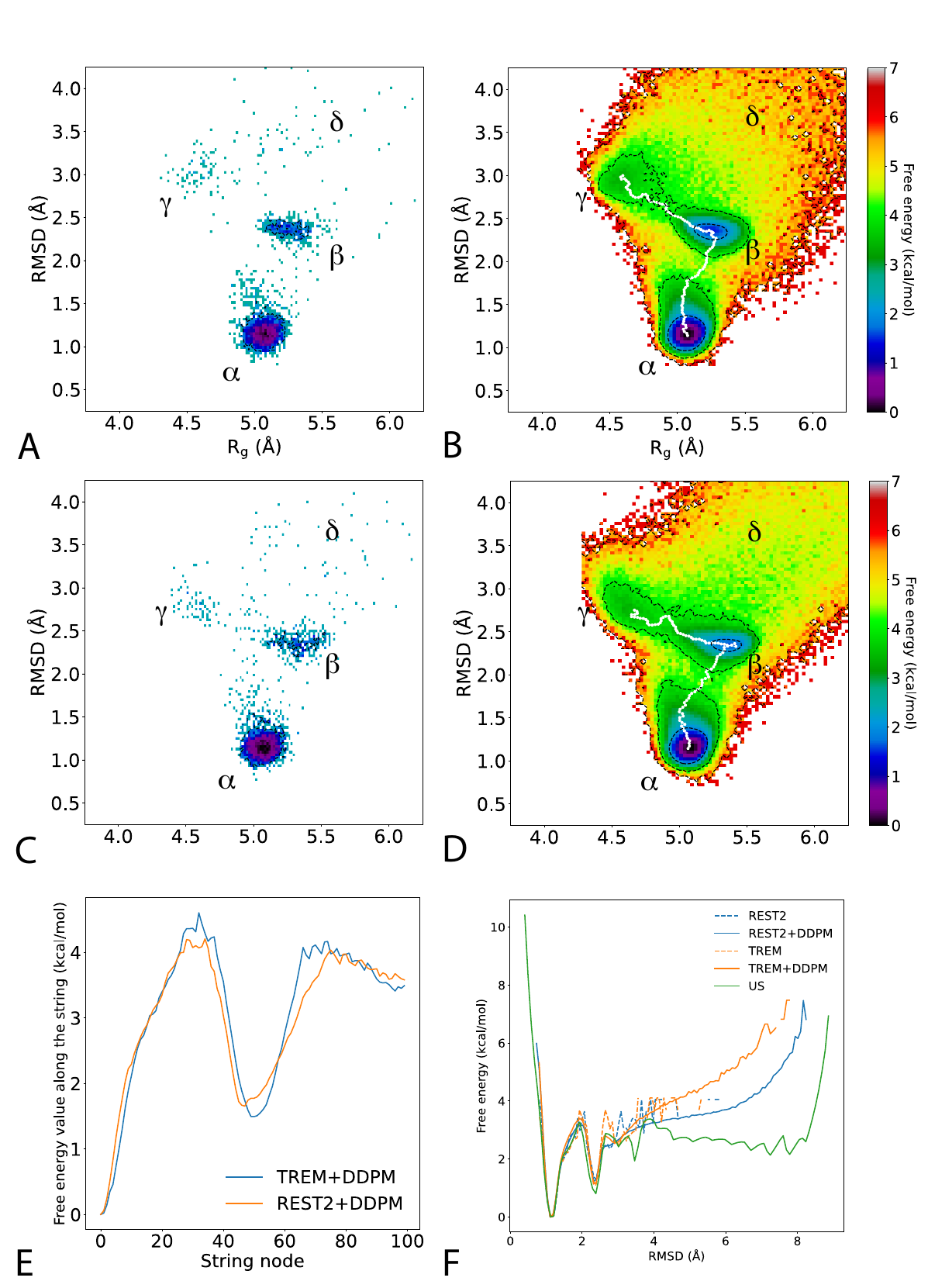} 
\caption{Comparison of two-dimensional free-energy surfaces computed along RMSD and $R_{g}$, obtained from TREM (A),  TREM refined by DDPM (B), REST2 (C), and REST2 refined by DDPM (D), as well as the free-energy values along the minimum free-energy path as predicted by DDPM (E). White lines on (B) and (D) show the minimum free-energy path. (F) One-dimensional free energy profile computed along the RMSD direction, comparing the training data from TREM and REST2 (dashed lines, orange and blue, respectively) with the DDPM predictions (solid lines, orange and blue, respectively), and Umbrella Sampling as a benchmark (solid green line). Representative structures of basins $\alpha$, $\beta$, $\gamma$, and $\delta$ are shown in Figure~S1.}
\label{fig:CLN_comp_REST2_TREM}
\end{figure}

By projecting the probability distributions estimated from the unscaled replica along (Figure~\ref{fig:CLN_comp_REST2_TREM}A and C), it is clear that both TREM and REST2 achieve comparable sampling performances. Key free-energy minima are identified, corresponding to radically different conformations (a few examples numbered $\alpha$ to $\gamma$ are shown in Figure~S1). We observe that while the basins corresponding to regions $\alpha$ and $\beta$ are well sampled, there are significantly fewer samples in $\gamma$ and $\delta$, and the transition regions are not well refined. As a consequence, once cannot clearly estimate the minimum free-energy paths and barriers connecting the apparent basins, in particular $\alpha$, $\beta$ and $\gamma$ on this two-dimensional surface.

When post-processing the results of the simulations with DDPM, and then use it to generate new samples for the unscaled replica conditions, much more converged free-energy surfaces can be recovered (Figure~\ref{fig:CLN_comp_REST2_TREM}B and D). In particular, the transition states regions connection $\alpha$, $\beta$ and $\gamma$ are now fully resolved. These are extrapolated from the model based on higher temperature/lower potential simulations learned from the model and extrapolated in the unperturbed replica. Indeed, the corresponding areas appear sampled when climbing the replica ladder (Figure~S2). 

In order to make a more quantitative comparison between the two approaches, we determined the minimum free-energy path connecting $\alpha$ to $\beta$ to $\gamma$ from the DDPM-TREM and DDPM-REST2 FES at 300~K. As shown in Figure~\ref{fig:CLN_comp_REST2_TREM}E, we do not find any statistically-relevant difference between both approaches. In particular, the relative free-energy differences between the metastable basins as well as the barrier heights are very similar. 

By defining regions corresponding to these three basins (see Methods) we can also estimate the total free-energy differences between $\alpha$ and $\beta$, and $\beta$ and $\gamma$ respectively. As shown in Table~\ref{tab:DG}, all approaches lead to similar results. In particular, the free-energy differences between the different basins was already well described without the use of higher replicas and DDPM. However, and quite notably, the comparison between Figure~\ref{fig:CLN_comp_REST2_TREM}A and B, and Figure~\ref{fig:CLN_comp_REST2_TREM}C and D, makes it clear that DDPM allows to resolve free-energy barriers on the 2D FES that are otherwise not well quantified by looking at the unperturbed replica only, both for TREM and REST2 trajectories.

\begin{table*}[h!]
\centering
\begin{tabular}{ccccc}
\hline
Free-energy difference & TREM & DDPM-TREM & REST2 & DDPM-REST2 \\
\hline
\hline

$F_\beta - F_\alpha $(kcal/mol)  & 1.11 $\pm$ 0.34 & 1.05 $\pm$ 0.33 & 1.23 $\pm$ 0.24 & 1.10 $\pm$ 0.24 \\
$F_\gamma - F_\beta $(kcal/mol) & 0.81 $\pm$ 0.36 & 0.72 $\pm$ 0.12& 0.64 $\pm$ 0.30 & 0.81 $\pm$ 0.50\\
\end{tabular}
\caption{Total free energy differences between basins $\alpha$ , $\beta$ and $\gamma$ computed from TREM, DDPM-TREM, REST2 and DDPM-REST2. Error bars for the TREM and REST2 simulations were obtained by calculating standard deviations across blocks of 100 ns, and those for DDPM were obtained by training the model on these different blocks.}
\label{tab:DG} 
\end{table*}

Although there is good agreement between the two replica exchange strategies combined with DDPM, there is no guarantee that they actually converge toward the actual ground-truth reference of the associated conformational change. Having determined from 2D-projections that the key bottleneck to conformational exploration was (in the chosen CV-space) transitions along the RMSD to the crystal structure, we also performed US simulations along this coordinate to estimate a ground-truth 1D FES to which we could compare the results of replica exchange simulations augmented with DDPM. We stress that these data are completely unrelated to the (replica-exchange) data used to train the DDPM model.

For RMSD values below 3~\AA, there is good agreement between the DDPM predictions trained on both TREM and REST2, and the umbrella sampling simulations. We note that in the one-dimensional projection, the incomplete and sparse description of barriers, visible in Figure~\ref{fig:CLN_comp_REST2_TREM}A and C, is much less problematic, and both REST2 and TREM readily provide reasonable estimates.

For RMSD larger than 3~\AA, the agreement between raw data from replica-exchange, DDPM, and US, is less accurate. This is due to the fact that high RMSD values correspond to a large number of possible unfolded, disordered conformations of CLN025 that are difficult to sample in all approaches (US simulations are also unlikely to be converged in that region). In particular, this indicates that the sampling provided by the replicas is not fully converged in this region of phase space. Indeed, sampling is pretty sparse and noisy in these high-RMSD regions, and even so for the highest temperature replica (Figure~S2). 

\subsection{Comparison with long, unbiased MD trajectories}

As an additional comparison with "ground-truth'' reference data, we repeated our simulations as well as DDPM training and generation for CLN025 in water under force-field and thermodynamic conditions matching those of a 106-$\mu$s trajectory from D.~E.~Shaw Research. As shown in Figure~S3, DDPM substantially improves the sampling of the 2D free-energy surface (FES). Notably, the higher temperature of these simulations (corresponding to the melting temperature of CLN025) together with this force field leads to considerably greater conformational plasticity even prior to applying DDPM, relative to the data presented above with a different force field at the more standard 300~K. Remarkably, DDPM yields more accurate estimates of the free-energy difference between the folded (RMSD $<2$~\AA) and unfolded (RMSD $>2$~\AA) basins compared to the raw REST2 results: $\Delta G = 0.83$~kcal/mol for the unbiased trajectory, $1.34 \pm 0.24$~kcal/mol for REST2, and $0.87 \pm 0.14$~kcal/mol for DDPM-REST2.

\subsection{Importance of exchanges for reliable reference data}

The fact that DDPM results remain sensitive to the quality of sampling in the replica exchange simulations is further illustrated by studying the impact of exchanges among replicas on the obtained results. Strikingly, it has been argued in the original DDPM-TREM formulation\cite{wang2022data}, as well as in subsequent work\cite{doi:10.1073/pnas.2321971121}, that DDPM could perform well even if exchanges were vanishingly small (smaller than a few \%, or even non existent). 

\begin{figure}[h!]
\centering
\includegraphics[width=0.7\linewidth]{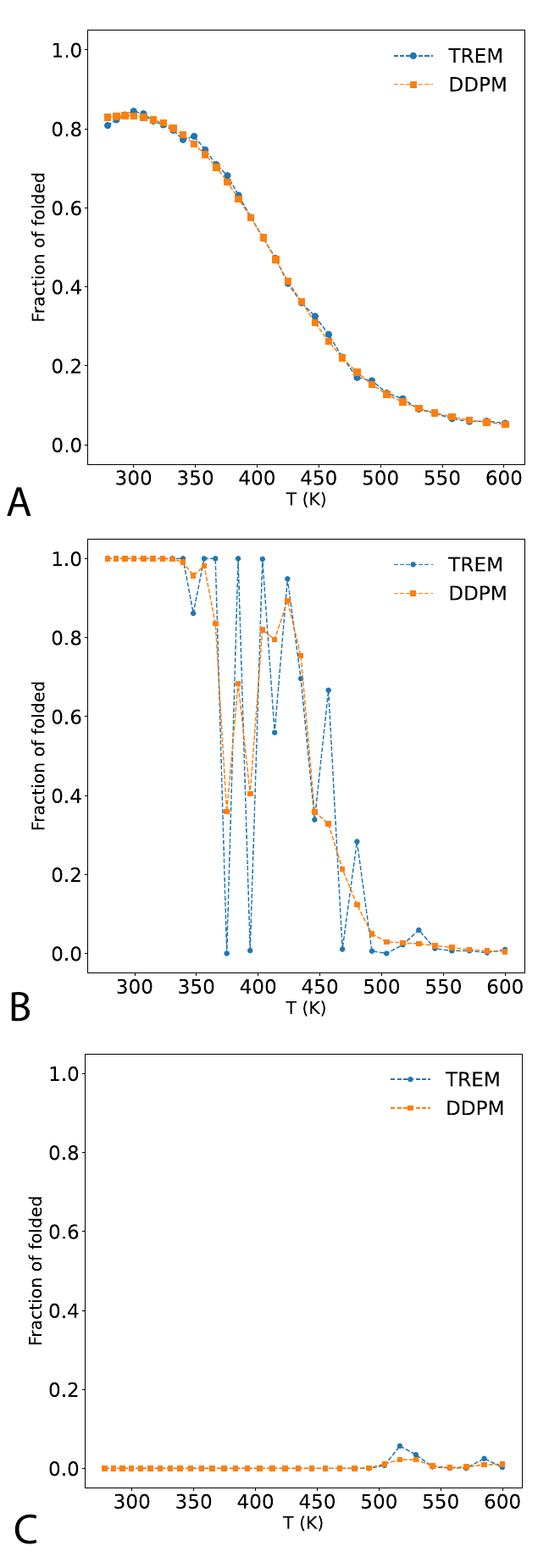} 
\caption{Melting curves showing the fraction of folded protein computed from TREM simulations (in orange) (A) with exchanges between replicas, (B) without exchange between replicas, with the same folded initial structure for all replicas, (C) without exchange between replicas, with the same unfolded initial structure for all replicas. The prediction of DDPM trained using these TREM simulations is shown in blue.}
\label{fig:fusion_exch}
\end{figure}

To do so, we calculated melting curves, which represent the fraction of folded structures (defined as RMSD smaller than 2~\AA) as a function of temperature. These curves were extracted from TREM simulations performed in three different ways: with exchanges (Figure~\ref{fig:fusion_exch}A), without exchanges and starting from a folded structure (Figure~\ref{fig:fusion_exch}B), and without exchanges and starting from an unfolded structure (Figure~\ref{fig:fusion_exch} C). 

As seen in these plots, exchanges are crucial for converging the melting curve, which is a measure of the convergence of the thermodynamic quantities, and they significantly impact DDPM predictions. For instance, DDPM trained on TREM without exchanges, starting with unfolded structures, cannot generate folded structures because no folding is observed in the TREM without exchanges. Starting from a folded structure, TREM simulations can generate unfolded structures in the high-temperature replicas, but the melting curves are very noisy and DDPM alone does not improve the sampling. 


The main difference with previous work is that (i) here exchanges are crucial to drive chignolin folding/unfolding, whereas some of the systems investigated previously did not exhibit folded/unfolded states\cite{wang2022data}; and (ii) in other more complex cases\cite{doi:10.1073/pnas.2321971121}, configurations are mixed across the replica ladder and periodically updated, therefore mitigating the effects of small exchange rates.

This highlights the advantage of using REST2 for training DDPM, as REST2 requires much fewer replicas compared to TREM. Regarding computational cost, the training of DDPM and data generation for REST2 and TREM in these examples were similarly affordable from a computational perspective (less than a dozen gpu.hrs). The gain in computational efficiency comes from the better scaling of REST2 with system size: we needed only 12 replicas with REST2, compared to 32 with TREM, achieving better exchange rates with REST2 (0.51) than with TREM (0.13) and yielding comparable sampling at the reference temperature (Figure \ref{fig:CLN_comp_REST2_TREM}A and C). This resulted in a 2.7-fold reduction in computational cost for REST2 compared to TREM for the same sampling efficiency. The gain would be even more obvious for a larger biomolecule that would require many more replicas in TREM.    

\subsection{Resolving large conformational changes with high activation barriers}

Having shown that the combination of DDPM and Hamiltonian replica exchange can greatly enhance exploration of the free-energy surface of a model, mini-protein system, we now turn toward the greater challenge posed by a larger protein exhibiting complex and hard-to-sample conformational changes. An additional requirement is to have reference simulation data about this conformational landscape as "ground-truth" to which we could compare our own results. 

We thus chose to focus on the loop opening/closing motion in the phosphatase PTP1B. PTP1B is a well-studied member of the Protein Tyrosine Phosphatase (PTP) superfamily, which is involved in multiple cellular processes such as glucose uptake, and proliferation \cite{zhang2001protein}. It is also a drug target for several diseases, including diabetes and cancer \cite{feldhammer2013ptp1b}.


\begin{figure*}[h!]
\centering
\includegraphics[width=\textwidth]{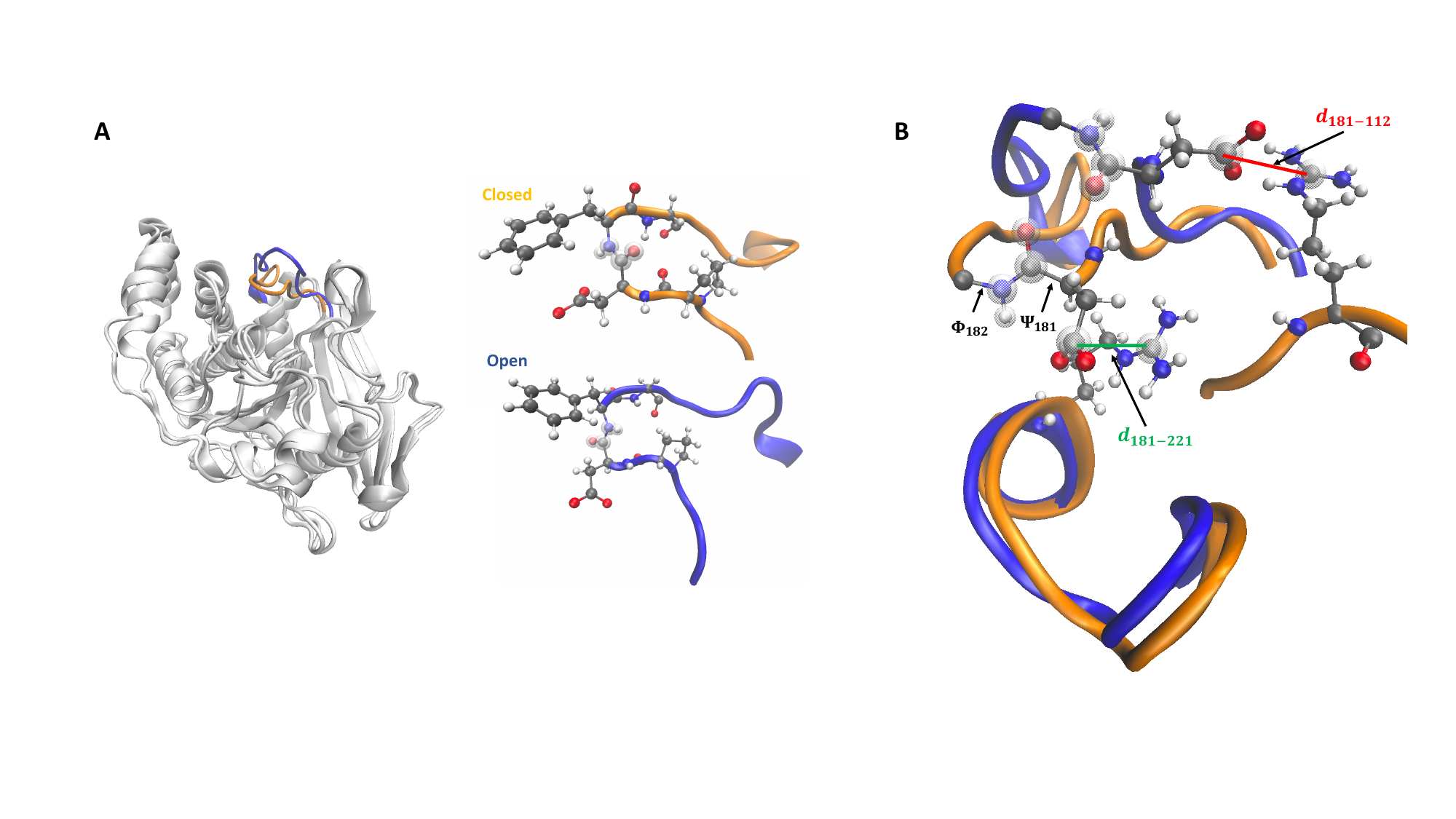} 
\caption{Superimposed closed and open conformations of PTP1B. A focus on the WPD loop is shown on the right, for the closed (blue) and the open (orange) conformations. The two dihedral angles used to distinguish the closed and open conformation are highlighted (A). The four collective variables used to train the DDPM, represented on the superimposed closed and open conformations of PTP1B. The two dihedral angles $\Psi_{181}$ and $\Phi_{182}$ are shown with an arrow, the distance between residue 181 and residue 221 is drawn in green, and the distance between residue 181 and residue 112 is drawn in red (B).}
\label{fig:PTP1B_loops}
\end{figure*}

Before we focus on our results, we now provide some context about this enzyme (Figure~\ref{fig:PTP1B_loops}A). PTP1B catalyzes the hydrolysis of phosphorylated tyrosine in a two-step process involving a thiophosphate enzyme intermediate. The first step consists of a nucleophilic attack by a cysteine residue on the phosphorylated tyrosine, breaking the phosphate bond. The second step involves the hydrolysis of the intermediate through a nucleophilic attack by a water molecule. Both steps require the assistance of an aspartic residue (Asp181) which acts as a proton donor and acceptor during the reaction. Asp181 belongs to the WPD-loop, a flexible loop comprising a dozen residues (177 to 188) that exists in two conformations: the hydrolysis-incompetent open state, and the hydrolysis-competent closed state (Figure~\ref{fig:PTP1B_loops}B). In the open state, Asp181 forms a salt-bridge with arginine Arg112, while in the closed state, Asp181 rotates to form a salt-bridge with Arg221, allowing Asp181 to get closer to the catalytic site for proton transfers. 

NMR studies have shown that the dynamics of the WPD-loop is the rate-limiting step of the reaction \cite{whittier2013conformational}, with a rate constant of $k_{open \rightarrow closed} = 22$ s$^{-1}$ for the open-to-closed transition, and $k_{closed \rightarrow open} = 890$ s$^{-1}$ for the closed-to-open transition, yielding an equilibrium constant $K = k_{closed \rightarrow open}/k_{open \rightarrow closed} = 40$ in favor of the open conformation. Computational studies \cite{shen2021single} have highlighted the effect of point mutations on the WPD loop, affecting its dynamics and enzymatic activity. A recent paper \cite{zinovjev2024activation}, using molecular dynamics and the adaptive string method \cite{zinovjev2017adaptive}, elucidated the mechanism and the rate-limiting step of the WPD loop dynamics, corresponding to the rotation of the peptide group involving residues Asp181 and Phe182.  The authors identified the key reaction coordinates for this process (Figure~\ref{fig:PTP1B_loops}B): two torsional angles $\psi_{181}$ and $\phi_{182}$ to describe the rotation, and two distances Asp181C$\gamma$-Arg112C$\zeta$ ($d_{181-112}$) and Asp181C$\gamma$-Arg221C$\zeta$ ($d_{181-221}$) to track the salt-bridges formed and broken in the closed and open conformations. They also computed the free-energy profile associated with the transition from closed to open, yielding a free-energy barrier of approximately 12 kcal/mol, and a free-energy difference between closed and open states close to the experimental value of 2.3 kcal/mol in favor of the open conformation \cite{whittier2013conformational}.

For such a large system, TREM simulations become impractical, because the balance between the extent of the temperature ladder and good exchange rates would require an unreasonable amount of replicas. Instead, REST2 is less sensitive to the system size and was used here with 24 replicas with scaling factors ranging from 1.034 to 0.5. 

\begin{figure*}[h!]
\centering
\includegraphics[width=0.8\textwidth]{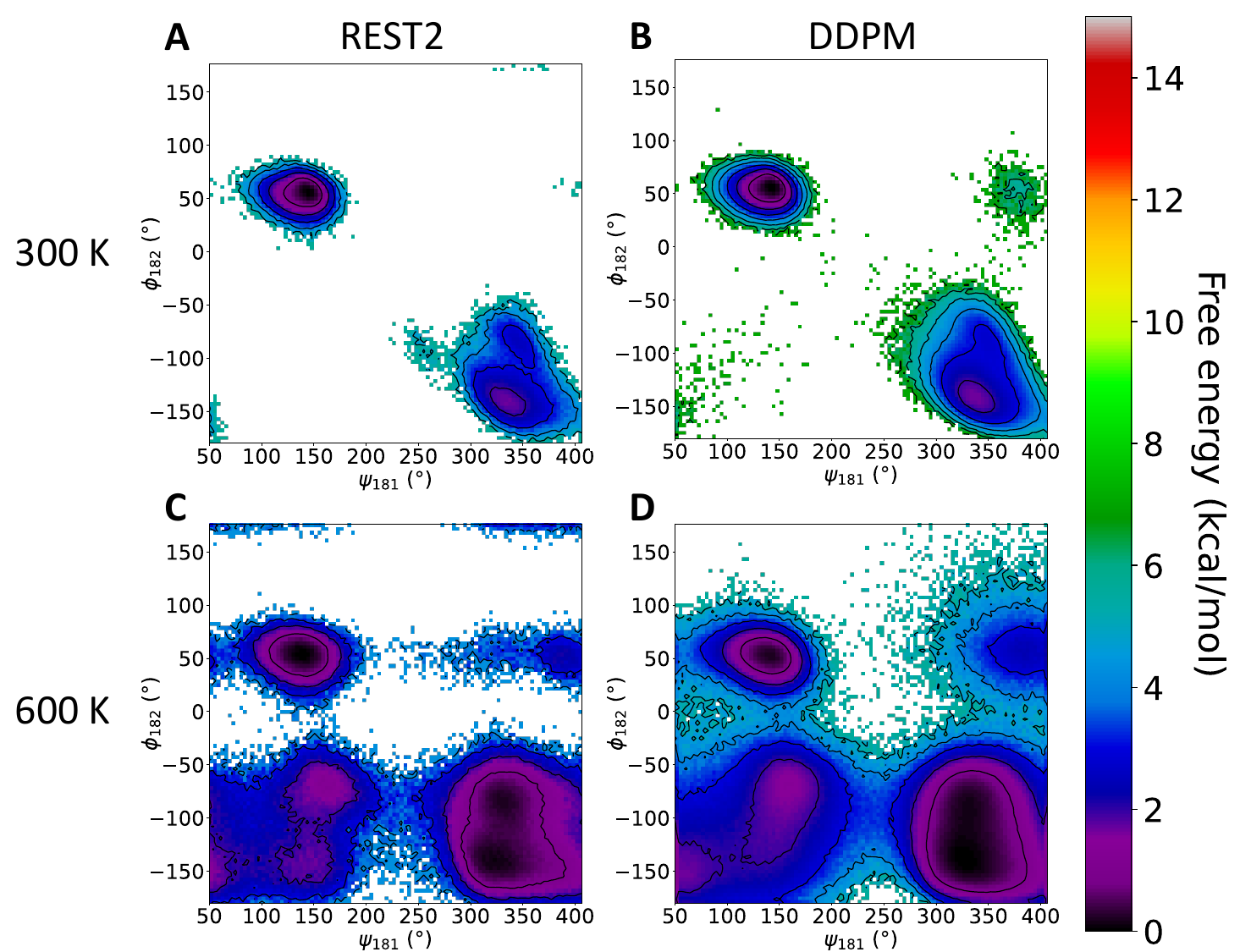} 
\caption{Free-energy surfaces computed along the two torsional angles $\psi_{181}$ and $\phi_{182}$. (A) REST2 simulations for the reference temperature replica. (B) DDPM for the reference temperature replica. (C) REST2 simulations for the highest temperature replica. (D) DDPM for the highest temperature replica.}
\label{fig:PTP1B_no_bias}
\end{figure*}

Starting first from a closed conformation (we provide further below the comparison with simulations starting from the open conformation), we propagated the simulations for 600 ns and analyzed the converged part of the trajectories. We computed the free-energy surface along the two torsional angles $\psi_{181}$ and $\phi_{182}$ whose changes allow the rotation of the Asp181-Phe182 peptide group, which have been shown to be responsible for the high free-energy barrier characterizing the transition \cite{zinovjev2024activation}. The FES (Figure~\ref{fig:PTP1B_no_bias}A) is characterized by two basins, corresponding to the closed (upper left) and open (bottom right) conformations; no transition path is sampled to go from the closed to the open conformation. In addition, the free-energy difference between the two basins is in favor of the closed conformation, at odds with previous investigations and experimental results (Table~\ref{tab:DG_PTP1B}).

\begin{table*}[h!]
\centering
\begin{tabular}{cccc}
\hline
Method & CV-bias  & Starting conformation & $ F_{\text{open}}-F_{\text{closed}}$ (kcal/mol) \\
\hline
\hline
REST2 & no & closed &  0.70 $\pm$ 0.24 \\
DDPM-REST2 & no & closed &  0.87 $\pm$ 0.25 \\
REST2 & yes & closed & $ -0.80$ $\pm$ 1.08 \\
DDPM-REST2 & yes & closed &  $-0.92$ $\pm$ 1.05  \\
DDPM-REST2 & yes & open &   $-1.75$ $\pm$ 1.26   \\

\end{tabular}
\caption{Free energy differences between closed  and open conformations, computed from REST2 starting closed, DDPM-REST2, REST2+CV-bias starting closed, REST2+CV-bias+DDPM starting closed, and REST2+DDPM+CV-biasias starting open. The experimental value for $ F_{\text{open}}-F_{\text{closed}}$ is $-2.3$~kcal/mol\cite{whittier2013conformational}. Error bars for the unbiased and biased REST2 simulations were obtained by calculating standard deviations across blocks of 100 ns and those for DDPM were obtained by training the model on these different blocks.}
\label{tab:DG_PTP1B}
\end{table*}

We then trained a DDPM following the procedure explained in the previous section, learning the joint probability distribution along the four collective variables that were shown to be relevant for the open/close process ($\psi_{181}$, $\phi_{182}$, $d_{181-112}$, $d_{181-221}$) and the rescaled potential energy. DDPM is able to generate new samples in the transition state region at the reference temperature (Figure~ \ref{fig:PTP1B_no_bias}B), to better resolve the open conformation basin (bottom right corner) and to explore a new conformational basin located around (360$^{\circ}$, 50$^{\circ}$). However,  it is not enough to resolve the transition and obtain a converged estimate of the free-energy barrier separating the two basins. Moreover, DDPM does not solve the issue of the relative stabilization of the closed conformation (Table~\ref{tab:DG_PTP1B}).  Interestingly, for the highest temperature replica, where the free-energy barrier is expected to be lower (Figure~ \ref{fig:PTP1B_no_bias}C), DDPM is able to resolve the barrier (Figure~ \ref{fig:PTP1B_no_bias}D), which is not the case from the raw data from this replica (Figure~ \ref{fig:PTP1B_no_bias}C). 


Given the inability of the reference-temperature, unperturbed replica to effectively sample the transition state region between the two basins (while the most perturbed replica more uniformly explores the $\psi_{181}$ and $\phi_{182}$ dihedral angles, previously identified as the slow collective variables in loop closing/opening), we leveraged this probability distribution to bias exploration across all replicas in a new set of REST2 simulations. 

\begin{figure}[t!]
\centering
\includegraphics[width=0.8\linewidth]{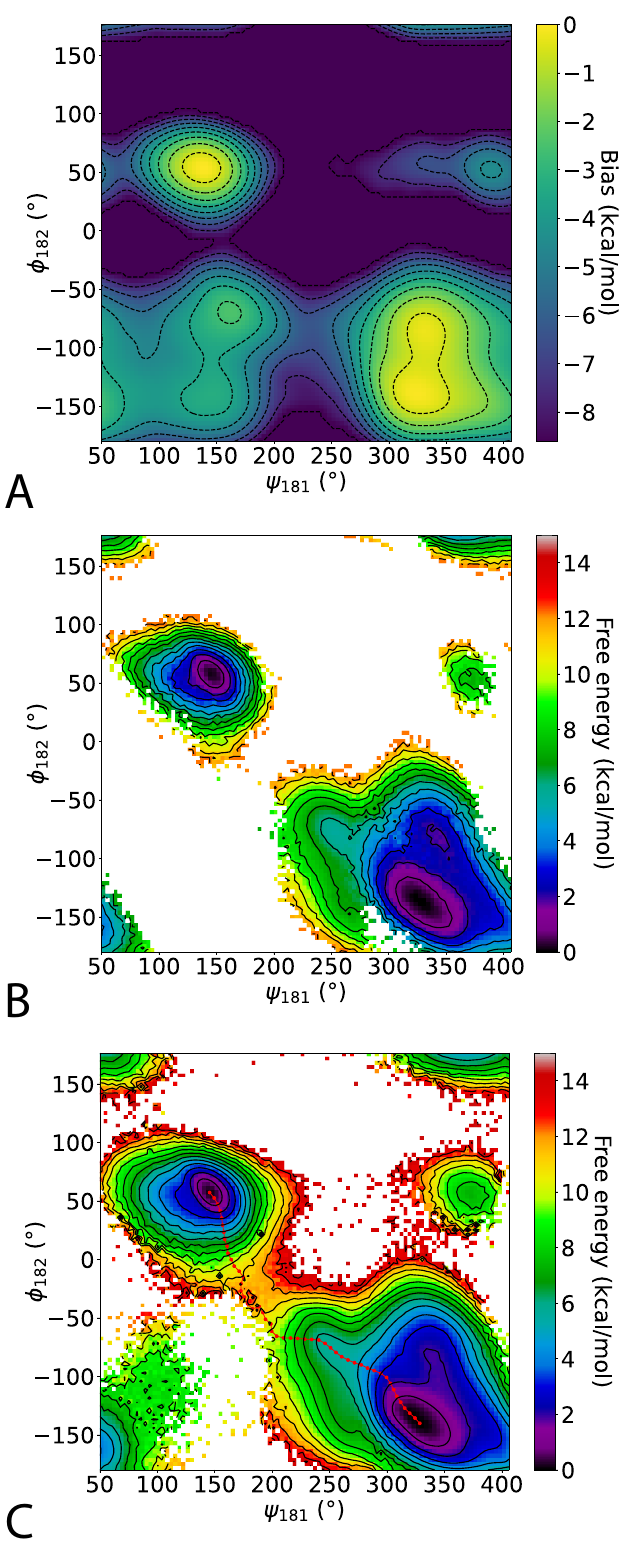} 
\caption{Prediction from DDPM trained on biased REST2 simulations initiated from the closed conformation.  (A) Two-dimensional bias applied to the reference temperature replica, computed following equation \ref{eq:bias}. (B) Free-energy surface obtained from the biased REST2 simulation, analyzing the reference temperature replica. (C) Prediction from DDPM trained with the biased REST2 simulations. DDPM is able to predict a second transition path, unseen in both the biased REST2 simulations and the KDE.}
\label{fig:PTP1B_bias}
\end{figure}

Specifically, we applied the following bias to the $i$-th replica:  

\begin{equation}
    V_{i}\left(\psi_{181},\phi_{182}\right) = \frac{\lambda_{i}}{\lambda_{high}} \times RT\ln\left(P_{high}^{KDE}(\psi_{181},\phi_{182})\right)
    \label{eq:bias}
\end{equation}

where $V_{i}\left(\psi_{181},\phi_{182}\right)$ is the bias applied along the torsional angles $\psi_{181}$ and $\phi_{182}$, $R$ is the gas constant, $T$ is the reference temperature (300 K), and $P_{high}^{KDE}(\psi_{181},\phi_{182})$ is the joint probability distribution of these angles, estimated using kernel density estimation (KDE) \cite{davis2011remarks,parzen1962estimation} for the high-temperature replica. The bias for each replica (see Figure~ \ref{fig:PTP1B_bias}A for the bias applied to the reference replica) is thus the negative of the free-energy profile computed along $\psi_{181}$ and $\phi_{182}$ at the highest temperature, rescaled by $\frac{\lambda_{i}}{\lambda_{high}}$ to account for the higher potential energy barriers at lower temperatures due to the REST2 rescaling scheme.  

Short REST2 simulations were then performed using a similar setup as the one described before but incorporating these biases, in the spirit of importance sampling. The presence of these biases significantly improved the exploration of the FES along $\psi_{181}$ and $\phi_{182}$. However, the reference replica still failed to fully sample the transition regions between the open and closed basins (Figure~\ref{fig:PTP1B_bias}B). By retraining and applying DDPM on these data, we obtained a FES at the reference temperature that reveals a clear transition pathway with an estimated barrier of approximately 11-12 kcal/mol (Figure~\ref{fig:PTP1B_bias}C, pathway in red). Notably, a KDE-based FES constructed from the reference replica alone yielded inaccurate transition barrier estimates, significantly exceeding those obtained from DDPM (14-15 kcal/mol, Figure~S4). While  methodological differences (as well as minor differences between the employed forcefields) between this work and the previous reference study\cite{zinovjev2024activation} could explain minor discrepancies, especially since the previous path was computed in a four-dimensional space incorporating $\psi_{181}$, $\phi_{182}$, $d_{181-112}$, and $d_{181-221}$, it is remarkable that both approaches converged to a nearly identical barrier of 11-12 kcal/mol\cite{zinovjev2024activation}.  


We note that in addition to the barrier, the biased REST2 simulation now provide better estimates of the free-energy difference between the open and closed conformations, in agreement with previous (unrelated) computational studies (Table~\ref{tab:DG_PTP1B}). We stress again that this previous data has obviously not been used to train the DDPM model. 

We conducted a committor analysis along this trajectory (Figure~S5) to confirm that the found reaction pathway actually represents a true transition path (Figure~S5). In particular, trajectories initiated around the corresponding transition state (initial configurations were selected along the REST2 trajectories) fall in either the closed or open basin, with a ratio close to 50:50 at the transition state location (Figure~S5). 

Additionally, while the transition region is not perfectly resolved, an alternative pathway with a similar barrier in the direction of decreasing $\psi_{181}$ values is visible in Figure~\ref{fig:PTP1B_bias}C,  passing through a transition station located around (120$^{\circ}$, -50$^{\circ}$). While it is not perfectly resolved starting from the closed state, it’s quite unambiguous when starting from the open conformation (see below). In contrast to the other pathway, a committor analysis reveals that a fraction of trajectories initiated along this pathway fail to reach either the open or closed states, instead becoming trapped in an intermediate region of the FES. This suggests that this alternative pathway may involve additional slow-relaxing variables, incurring extra free-energy and/or kinetic costs.  

\begin{figure}[t!]
\centering
\includegraphics[width=\linewidth]{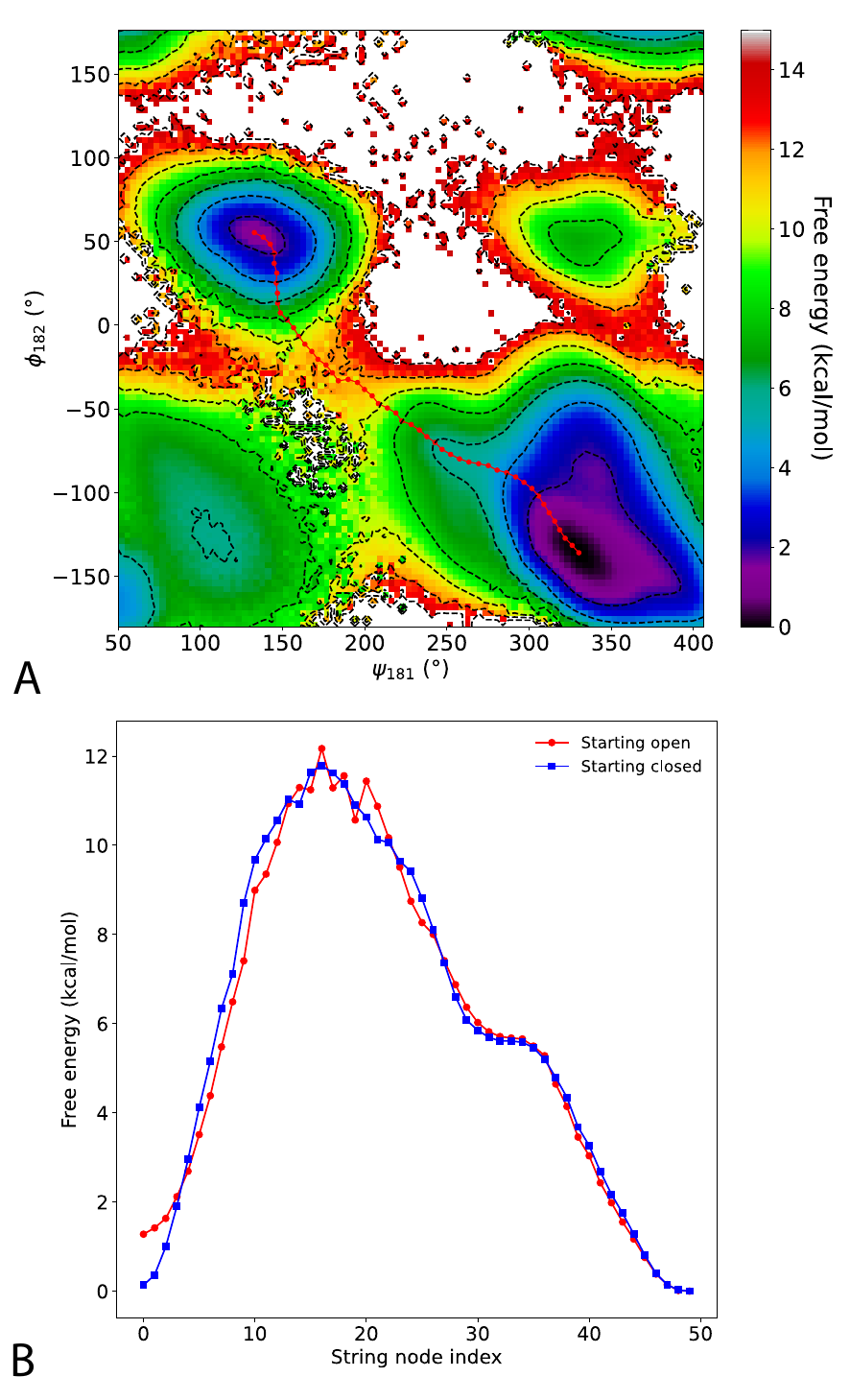} 
\caption{Prediction from DDPM trained on biased REST2 simulations initiated from the open conformation.
(A) Free energy surface along the $\psi$ and $\phi$ dihedral angles. (B) One-dimensional projection of the two-dimensional energy profile along the transition path, for simulations initiated from the closed conformation (blue) and the open conformation (red).}
\label{fig:PTP1B_bias2}
\end{figure}

The combined use of REST2, DDPM, and importance sampling provides robust estimates of the FES along what are indeed slow modes of the loop motion. We finally verified that the replica-exchange simulations were converged, in that sense that they were not sensitive to the starting structure. We thus repeated the simulations using the same biases but starting from the open loop conformation. Applying the same protocol, we identified a similar transition pathway, demonstrating that our method reliably captures the thermodynamic and mechanistic characteristics of the open/close motion (Figure~\ref{fig:PTP1B_bias2}A). In particular, the 1D projection of the projection path reveals very identical barriers (Figure~\ref{fig:PTP1B_bias2}B), with minor differences < 1 kcal/mol regarding the free-energy difference between the two conformations (Figure~\ref{fig:PTP1B_bias2}B and Table~\ref{tab:DG_PTP1B}).

\section*{Conclusion}

In this work, we show how conformational phase space exploration in Hamiltonian replica exchange simulations could better be exploited by a stochastic generative model (DDPM) to enhance the thermodynamic and kinetic mapping of the unperturbed system. While our work is very close in spirit from previous applications of DDPMs to temperature replica exchange simulations pioneered by the Tiwary group \cite{wang2022data,doi:10.1073/pnas.2321971121,lee25}, we propose here two key improvements. The first one is the extension to Hamiltonian replica exchange (HREX, here with the solute tempering approach REST2) simulations by treating the potential energy (or a fraction of it) as a random fluctuating variable in the framework of the diffusive model. This presents the distinctive advantage of enabling the study of larger systems compared to TREM as conformational exploration and exchange amongst the replicas scales much better with system size. For systems for which both TREM and HREX simulations can reliably converge, we demonstrate how DDPM can greatly enhance the determination of FES while yielding identical results for the two generalized ensemble schemes. 

Because generative models remain inherently limited by the quality of the data they have been trained on, we also realized that DDPM alone could not allow for the accurate determination of significant conformational barriers separating different (sampled) conformational basins. These states are only very transiently sampled during the conformational exploration of the replica-exchange simulations, which does not allow for successful DDPM training. We hereby propose a second improvement of the method, which consists in a combination of importance sampling with replica exchange and DDPM that could be iterated to gradually improve the description of the free-energy barrier(s). We apply this methodology for a complex protein system and demonstrate that a loop opening/closing pathway consistent with a previous investigation using an adaptive string method. 

An interesting aspect of this importance sampling approach, if the bias is first determined using DDPM itself, and provided some or all slow modes of the targeted conformational change are known, is that this could serve as an additional safeguard to verify that the diffusive model is not hallucinating. There are easy additional checks that can be easily performed once the position of a barrier has been identified on the FES, such as committor analyses, to confirm that these indeed correspond to transition pathways. 

A critical point for this second application to realistic systems is that the biased sampling requires to know, or to identify, the key slow collective variables of the system, which seems to go against the original spirit of generalized ensemble approaches that are CV-agnostic and allow to excite all the modes of the system. However, we note that this can actually be a distinct advantage. One could specifically target a known slow mode of the system while facilitating sampling along faster (but not necessarily fast enough to be fully sampled during plain unperturbed MD simulations) degrees of freedom. In a sense, it is very similar in spirit to hybrid schemes combining generalized ensemble and specific biasing approaches\cite{henin2022enhanced}. 

Here, we focused on two systems for which we knew relevant collective variables a priori, both for DDPM training and for biasing in cases the barrier remained inaccessible. We note however that DDPM has been shown to be effective even when using dozens of CVs\cite{wang2022data,doi:10.1073/pnas.2321971121,lee25}. Therefore, the identification of suitable reaction coordinates can be approached in two complementary ways. One option is to train the DDPM on an extended set of CVs, followed by the use of a machine learning-based technique\cite{annurev:/content/journals/10.1146/annurev-physchem-083122-125941}, or a non-ML alternative, to identify the system’s slow modes and transformation pathways. Alternatively, such techniques can first be employed to identify a relevant reaction coordinate, after which our scheme could be applied by biasing along this chosen CV while accelerating sampling along all other modes of the system. 

We conclude by stressing that although key limitations remain (sufficient sampling by the replica exchange, identification of reaction coordinates in more complex cases if one wants to determine accurate barriers), the training of the generative model and the subsequent generation of samples comes at a negligible computational cost compared to that of the replica exchange simulations. The overall computational gain stems from its ability to leverage information from all replicas to enhance the accuracy of the probability distribution, and consequently the free-energy surfaces, in the reference, unperturbed replica. As such, this provides the same information as much longer replica-exchange trajectories that would not mix information between the replicas during the post-processing and analysis phases. 

At worse, this therefore appears as a useful and easily-usable post-processing tool that can greatly enhance the thermodynamic interpretation and exploitation of replica exchange simulation data, but also of regular MD data, as recently shown. At best, and combined with specific targeting of relevant collective variables, this enables the determination of complex conformational pathways that are not necessarily easily determined using other computational techniques.

\section{Acknowledgements}

The research leading to these results has received funding from the European Research Council under the European Union's  Eighth Framework Program (H2020/2014-2020)/ERC Grant Agreement No. 757111 (G.S.). This work was also supported by the "Initiative d'Excellence" program from the French State (Grant "DYNAMO", ANR-11-LABX-0011-01 to GS). ZB was funded by a PhD Fellowship from the CFM Research Foundation. The simulations presented here benefited from access to the HPC resources of TGCC under the allocation A0150811005 granted by GENCI (Grand Equipement National de Calcul Intensif). 

\section{Conflicts of interest}
The authors have no competing interests to declare. 

\section{Data availability}
The DDPM-REST2  code is available on Github (stirnemann-group/rest2-ddpm), together with some examples. All inputs and representative simulation trajectories are available on Zenodo, doi: 10.5281/zenodo.17072659. 

\section{Supporting information}
Additional figures (representative chignolin conformations, exploration in low-potential replicas, comparison to long unbiased MD trajectories for CLN025, KDE-based free-energy surface reconstruction for PTP1B, committor analysis).


\begin{mcitethebibliography}{52}
\providecommand*\natexlab[1]{#1}
\providecommand*\mciteSetBstSublistMode[1]{}
\providecommand*\mciteSetBstMaxWidthForm[2]{}
\providecommand*\mciteBstWouldAddEndPuncttrue
  {\def\EndOfBibitem{\unskip.}}
\providecommand*\mciteBstWouldAddEndPunctfalse
  {\let\EndOfBibitem\relax}
\providecommand*\mciteSetBstMidEndSepPunct[3]{}
\providecommand*\mciteSetBstSublistLabelBeginEnd[3]{}
\providecommand*\EndOfBibitem{}
\mciteSetBstSublistMode{f}
\mciteSetBstMaxWidthForm{subitem}{(\alph{mcitesubitemcount})}
\mciteSetBstSublistLabelBeginEnd
  {\mcitemaxwidthsubitemform\space}
  {\relax}
  {\relax}

\bibitem[Tuckerman(2023)]{tuckerman2023statistical}
Tuckerman,~M.~E. \emph{Statistical Mechanics: Theory and Molecular Simulation};
  Oxford university press, \textbf{2023}\relax
\mciteBstWouldAddEndPuncttrue
\mciteSetBstMidEndSepPunct{\mcitedefaultmidpunct}
{\mcitedefaultendpunct}{\mcitedefaultseppunct}\relax
\EndOfBibitem
\bibitem[H{\'e}nin \latin{et~al.}(2022)H{\'e}nin, Leli{\`e}vre, Shirts,
  Valsson, and Delemotte]{henin2022enhanced}
H{\'e}nin,~J.; Leli{\`e}vre,~T.; Shirts,~M.~R.; Valsson,~O.; Delemotte,~L.
  Enhanced Sampling Methods for Molecular Dynamics Simulations. \emph{LiveCoMS} \textbf{2022}, \emph{4},
  1583\relax
\mciteBstWouldAddEndPuncttrue
\mciteSetBstMidEndSepPunct{\mcitedefaultmidpunct}
{\mcitedefaultendpunct}{\mcitedefaultseppunct}\relax
\EndOfBibitem
\bibitem[Torrie and Valleau(1977)Torrie, and Valleau]{torrie1977nonphysical}
Torrie,~G.~M.; Valleau,~J.~P. Nonphysical Sampling Distributions in Monte Carlo
  Free-Energy Estimation: Umbrella Sampling. \emph{J. Comput. Phys} \textbf{1977}, \emph{23}, 187--199\relax
\mciteBstWouldAddEndPuncttrue
\mciteSetBstMidEndSepPunct{\mcitedefaultmidpunct}
{\mcitedefaultendpunct}{\mcitedefaultseppunct}\relax
\EndOfBibitem
\bibitem[Laio and Parrinello(2002)Laio, and Parrinello]{laio2002escaping}
Laio,~A.; Parrinello,~M. Escaping Free-Energy Minima. \emph{Proc. Natl. Acad. Sci. U.S.A} \textbf{2002}, \emph{99}, 12562--12566\relax
\mciteBstWouldAddEndPuncttrue
\mciteSetBstMidEndSepPunct{\mcitedefaultmidpunct}
{\mcitedefaultendpunct}{\mcitedefaultseppunct}\relax
\EndOfBibitem
\bibitem[Barducci \latin{et~al.}(2008)Barducci, Bussi, and
  Parrinello]{barducci2008well}
Barducci,~A.; Bussi,~G.; Parrinello,~M. Well-Tempered Metadynamics: a Smoothly
  Converging and Tunable Free-Energy Method. \emph{Phys. Rev. Lett}
  \textbf{2008}, \emph{100}, 020603\relax
\mciteBstWouldAddEndPuncttrue
\mciteSetBstMidEndSepPunct{\mcitedefaultmidpunct}
{\mcitedefaultendpunct}{\mcitedefaultseppunct}\relax
\EndOfBibitem
\bibitem[Invernizzi and Parrinello(2020)Invernizzi, and
  Parrinello]{invernizzi2020rethinking}
Invernizzi,~M.; Parrinello,~M. Rethinking Metadynamics: from Bias Potentials to
  Probability Distributions. \emph{J. Phys. Chem. Lett}
  \textbf{2020}, \emph{11}, 2731--2736\relax
\mciteBstWouldAddEndPuncttrue
\mciteSetBstMidEndSepPunct{\mcitedefaultmidpunct}
{\mcitedefaultendpunct}{\mcitedefaultseppunct}\relax
\EndOfBibitem
\bibitem[Bussi and Laio(2020)Bussi, and Laio]{bussi2020using}
Bussi,~G.; Laio,~A. Using Metadynamics to Explore Complex Free-Energy
  Landscapes. \emph{Nat. Rev. Phys} \textbf{2020}, \emph{2},
  200--212\relax
\mciteBstWouldAddEndPuncttrue
\mciteSetBstMidEndSepPunct{\mcitedefaultmidpunct}
{\mcitedefaultendpunct}{\mcitedefaultseppunct}\relax
\EndOfBibitem
\bibitem[Rosso \latin{et~al.}(2002)Rosso, Min{\'a}ry, Zhu, and
  Tuckerman]{10.1063/1.1448491}
Rosso,~L.; Min{\'a}ry,~P.; Zhu,~Z.; Tuckerman,~M.~E. On the Use of the
  Adiabatic Molecular Dynamics Technique in the Calculation of Free Energy
  Profiles. \emph{J. Chem. Phys} \textbf{2002}, \emph{116},
  4389--4402\relax
\mciteBstWouldAddEndPuncttrue
\mciteSetBstMidEndSepPunct{\mcitedefaultmidpunct}
{\mcitedefaultendpunct}{\mcitedefaultseppunct}\relax
\EndOfBibitem
\bibitem[Maragliano and Vanden-Eijnden(2006)Maragliano, and
  Vanden-Eijnden]{MARAGLIANO2006168}
Maragliano,~L.; Vanden-Eijnden,~E. A Temperature Accelerated Method for
  Sampling Free Energy and Determining Reaction  Pathways in Rare Events
  Simulations. \emph{Chem. Phys. Lett} \textbf{2006}, \emph{426},
  168--175\relax
\mciteBstWouldAddEndPuncttrue
\mciteSetBstMidEndSepPunct{\mcitedefaultmidpunct}
{\mcitedefaultendpunct}{\mcitedefaultseppunct}\relax
\EndOfBibitem
\bibitem[Bolhuis \latin{et~al.}(2002)Bolhuis, Chandler, Dellago, and
  Geissler]{bolhuis2002transition}
Bolhuis,~P.~G.; Chandler,~D.; Dellago,~C.; Geissler,~P.~L. Transition path
  sampling: Throwing Ropes over Rough Mountain Passes, in the Dark.
  \emph{Annu. Rev. Phys. Chem} \textbf{2002}, \emph{53},
  291--318\relax
\mciteBstWouldAddEndPuncttrue
\mciteSetBstMidEndSepPunct{\mcitedefaultmidpunct}
{\mcitedefaultendpunct}{\mcitedefaultseppunct}\relax
\EndOfBibitem
\bibitem[Dellago \latin{et~al.}(1998)Dellago, Bolhuis, and
  Chandler]{dellago1998efficient}
Dellago,~C.; Bolhuis,~P.~G.; Chandler,~D. Efficient Transition Path Sampling:
  Application to Lennard-Jones Cluster Rearrangements. \emph{J. Chem. Phys} \textbf{1998}, \emph{108}, 9236--9245\relax
\mciteBstWouldAddEndPuncttrue
\mciteSetBstMidEndSepPunct{\mcitedefaultmidpunct}
{\mcitedefaultendpunct}{\mcitedefaultseppunct}\relax
\EndOfBibitem
\bibitem[Chong \latin{et~al.}(2017)Chong, Saglam, and Zuckerman]{CHONG201788}
Chong,~L.~T.; Saglam,~A.~S.; Zuckerman,~D.~M. Path-Sampling Strategies for
  Simulating Rare Events in Biomolecular Systems. \emph{Curr. Opin. Struct. Biol.} \textbf{2017}, \emph{43}, 88--94\relax
\mciteBstWouldAddEndPuncttrue
\mciteSetBstMidEndSepPunct{\mcitedefaultmidpunct}
{\mcitedefaultendpunct}{\mcitedefaultseppunct}\relax
\EndOfBibitem
\bibitem[van Erp(2007)]{PhysRevLett.98.268301}
van Erp,~T.~S. Reaction Rate Calculation by Parallel Path Swapping. \emph{Phys.
  Rev. Lett.} \textbf{2007}, \emph{98}, 268301\relax
\mciteBstWouldAddEndPuncttrue
\mciteSetBstMidEndSepPunct{\mcitedefaultmidpunct}
{\mcitedefaultendpunct}{\mcitedefaultseppunct}\relax
\EndOfBibitem
\bibitem[Zhang \latin{et~al.}(2024)Zhang, Baldauf, Roet, Lervik, and van
  Erp]{doi:10.1073/pnas.2318731121}
Zhang,~D.~T.; Baldauf,~L.; Roet,~S.; Lervik,~A.; van Erp,~T.~S. Highly
  Parallelizable Path Sampling with Minimal Rejections Using Asynchronous
  Replica Exchange and Infinite Swaps. \emph{Proc. Natl. Acad. Sci. USA} \textbf{2024}, \emph{121}, e2318731121\relax
\mciteBstWouldAddEndPuncttrue
\mciteSetBstMidEndSepPunct{\mcitedefaultmidpunct}
{\mcitedefaultendpunct}{\mcitedefaultseppunct}\relax
\EndOfBibitem

\bibitem[Sugita and Okamoto(1999)Sugita, and Okamoto]{sugita1999replica}
Sugita,~Y.; Okamoto,~Y. Replica-Exchange Molecular Dynamics Method for Protein
  folding. \emph{Chem. Phys. Lett.} \textbf{1999}, \emph{314},
  141--151\relax
\mciteBstWouldAddEndPuncttrue
\mciteSetBstMidEndSepPunct{\mcitedefaultmidpunct}
{\mcitedefaultendpunct}{\mcitedefaultseppunct}\relax
\EndOfBibitem
\bibitem[Wang \latin{et~al.}(2011)Wang, Friesner, and Berne]{wang2011replica}
Wang,~L.; Friesner,~R.~A.; Berne,~B. Replica Exchange with Solute Scaling: a
  More Efficient Version of Replica Exchange with Solute Tempering (REST2).
  \emph{J. Phys. Chem. B} \textbf{2011}, \emph{115},
  9431--9438\relax
\mciteBstWouldAddEndPuncttrue
\mciteSetBstMidEndSepPunct{\mcitedefaultmidpunct}
{\mcitedefaultendpunct}{\mcitedefaultseppunct}\relax
\EndOfBibitem
\bibitem[Marinari and Parisi(1992)Marinari, and Parisi]{marinari1992simulated}
Marinari,~E.; Parisi,~G. Simulated Tempering: a New Monte Carlo Scheme.
  \emph{EPL} \textbf{1992}, \emph{19}, 451\relax
\mciteBstWouldAddEndPuncttrue
\mciteSetBstMidEndSepPunct{\mcitedefaultmidpunct}
{\mcitedefaultendpunct}{\mcitedefaultseppunct}\relax
\EndOfBibitem
\bibitem[Hamelberg \latin{et~al.}(2004)Hamelberg, Mongan, and
  McCammon]{10.1063/1.1755656}
Hamelberg,~D.; Mongan,~J.; McCammon,~J.~A. Accelerated Molecular Dynamics: A
  Promising and Efficient Simulation Method for Biomolecules. \emph{J. Chem. Phys} \textbf{2004}, \emph{120}, 11919--11929\relax
\mciteBstWouldAddEndPuncttrue
\mciteSetBstMidEndSepPunct{\mcitedefaultmidpunct}
{\mcitedefaultendpunct}{\mcitedefaultseppunct}\relax
\EndOfBibitem
\bibitem[Lao \latin{et~al.}(2024)Lao, O’Connor, and Huang]{rev2024}
Lao,~Y.; O’Connor,~M.; Huang,~X. Replica Exchange with Solute Tempering for
  Protein Conformational Sampling. \emph{ChemRxiv} \textbf{2024}, \relax
\mciteBstWouldAddEndPunctfalse
\mciteSetBstMidEndSepPunct{\mcitedefaultmidpunct}
{}{\mcitedefaultseppunct}\relax
\EndOfBibitem
\bibitem[Languin-Catto{\"e}n \latin{et~al.}(2023)Languin-Catto{\"e}n, Sterpone,
  and Stirnemann]{LANGUINCATTOEN20232744}
Languin-Catto{\"e}n,~O.; Sterpone,~F.; Stirnemann,~G. Binding Site Plasticity
  Regulation of the FimH Catch-bond Mechanism. \emph{Biophys. J}
  \textbf{2023}, \emph{122}, 2744--2756\relax
\mciteBstWouldAddEndPuncttrue
\mciteSetBstMidEndSepPunct{\mcitedefaultmidpunct}
{\mcitedefaultendpunct}{\mcitedefaultseppunct}\relax
\EndOfBibitem
\bibitem[R{\"o}der \latin{et~al.}(2019)R{\"o}der, Stirnemann, Dock-Bregeon,
  Wales, and Pasquali]{10.1093/nar/gkz1071}
R{\"o}der,~K.; Stirnemann,~G.; Dock-Bregeon,~A.-C.; Wales,~D.~J.; Pasquali,~S.
  Structural Transitions in the RNA 7SK 5' Hairpin and their Effect on HEXIM
  Binding. \emph{Nucleic Acids Res} \textbf{2019}, \emph{48},
  373--389\relax
\mciteBstWouldAddEndPuncttrue
\mciteSetBstMidEndSepPunct{\mcitedefaultmidpunct}
{\mcitedefaultendpunct}{\mcitedefaultseppunct}\relax
\EndOfBibitem
\bibitem[Gallicchio \latin{et~al.}(2005)Gallicchio, Andrec, Felts, and
  Levy]{gallicchio2005temperature}
Gallicchio,~E.; Andrec,~M.; Felts,~A.~K.; Levy,~R.~M. Temperature Weighted
  Histogram Analysis Method, Replica Exchange, and Transition Paths. \emph{J. Phys. Chem. B} \textbf{2005}, \emph{109}, 6722--6731\relax
\mciteBstWouldAddEndPuncttrue
\mciteSetBstMidEndSepPunct{\mcitedefaultmidpunct}
{\mcitedefaultendpunct}{\mcitedefaultseppunct}\relax
\EndOfBibitem
\bibitem[Shirts and Chodera(2008)Shirts, and Chodera]{shirts2008statistically}
Shirts,~M.~R.; Chodera,~J.~D. Statistically Optimal Analysis of Samples from
  Multiple Equilibrium States. \emph{J. Chem. Phys}
  \textbf{2008}, \emph{129}\relax
\mciteBstWouldAddEndPuncttrue
\mciteSetBstMidEndSepPunct{\mcitedefaultmidpunct}
{\mcitedefaultendpunct}{\mcitedefaultseppunct}\relax
\EndOfBibitem
\bibitem[Wang \latin{et~al.}(2022)Wang, Herron, and Tiwary]{wang2022data}
Wang,~Y.; Herron,~L.; Tiwary,~P. From Data to Noise to Data for Mixing Physics
  across Temperatures with Generative Artificial Intelligence.
  \emph{Proc. Natl. Acad. Sci. U.S.A} \textbf{2022},
  \emph{119}, e2203656119\relax
\mciteBstWouldAddEndPuncttrue
\mciteSetBstMidEndSepPunct{\mcitedefaultmidpunct}
{\mcitedefaultendpunct}{\mcitedefaultseppunct}\relax
\EndOfBibitem
\bibitem[Ho \latin{et~al.}(2020)Ho, Jain, and Abbeel]{ho2020denoising}
Ho,~J.; Jain,~A.; Abbeel,~P. Denoising Diffusion Probabilistic Models.
  \emph{Adv. Neural Inf. Process.} \textbf{2020},
  \emph{33}, 6840--6851\relax
\mciteBstWouldAddEndPuncttrue
\mciteSetBstMidEndSepPunct{\mcitedefaultmidpunct}
{\mcitedefaultendpunct}{\mcitedefaultseppunct}\relax
\EndOfBibitem
\bibitem[Mehdi \latin{et~al.}(2024)Mehdi, Smith, Herron, Zou, and
  Tiwary]{annurev:/content/journals/10.1146/annurev-physchem-083122-125941}
Mehdi,~S.; Smith,~Z.; Herron,~L.; Zou,~Z.; Tiwary,~P. Enhanced Sampling with
  Machine Learning. \emph{Annu. Rev. Phys. Chem} \textbf{2024},
  \emph{75}, 347--370\relax
\mciteBstWouldAddEndPuncttrue
\mciteSetBstMidEndSepPunct{\mcitedefaultmidpunct}
{\mcitedefaultendpunct}{\mcitedefaultseppunct}\relax
\EndOfBibitem
\bibitem[Song \latin{et~al.}(2021)Song, Sohl-Dickstein, Kingma, Kumar, Ermon,
  and Poole]{song2021scorebasedgenerativemodelingstochastic}
Song,~Y.; Sohl-Dickstein,~J.; Kingma,~D.~P.; Kumar,~A.; Ermon,~S.; Poole,~B.
  Score-Based Generative Modeling through Stochastic Differential Equations.
  2021; \url{https://arxiv.org/abs/2011.13456}\relax
\mciteBstWouldAddEndPuncttrue
\mciteSetBstMidEndSepPunct{\mcitedefaultmidpunct}
{\mcitedefaultendpunct}{\mcitedefaultseppunct}\relax
\EndOfBibitem
\bibitem[Herron \latin{et~al.}(2024)Herron, Mondal, Schneekloth, and
  Tiwary]{doi:10.1073/pnas.2321971121}
Herron,~L.; Mondal,~K.; Schneekloth,~J.~S.; Tiwary,~P. Inferring Phase
  Transitions and Critical Exponents from Limited Observations with
  Thermodynamic Maps. \emph{Proc. Natl. Acad. Sci. U.S.A}
  \textbf{2024}, \emph{121}, e2321971121\relax
\mciteBstWouldAddEndPuncttrue
\mciteSetBstMidEndSepPunct{\mcitedefaultmidpunct}
{\mcitedefaultendpunct}{\mcitedefaultseppunct}\relax
\EndOfBibitem
\bibitem[Bera and Mondal(2025)Bera, and Mondal]{Bera2025.01.16.633470}
Bera,~P.; Mondal,~J. Assessing Generative Diffusion Models for Enhanced
  Sampling of Folded and Disordered Protein States Across Scales and in
  All-atom Resolution. \emph{bioRxiv} \textbf{2025},
  \emph{2025.01.16.633470}\relax
\mciteBstWouldAddEndPuncttrue
\mciteSetBstMidEndSepPunct{\mcitedefaultmidpunct}
{\mcitedefaultendpunct}{\mcitedefaultseppunct}\relax
\EndOfBibitem
\bibitem[Stirnemann and Sterpone(2015)Stirnemann, and
  Sterpone]{stirnemann2015recovering}
Stirnemann,~G.; Sterpone,~F. Recovering Protein Thermal Stability Using
  All-Atom Hamiltonian Replica-Exchange Simulations in Explicit Solvent.
  \emph{J. Chem. Theory Comput} \textbf{2015}, \emph{11},
  5573--5577\relax
\mciteBstWouldAddEndPuncttrue
\mciteSetBstMidEndSepPunct{\mcitedefaultmidpunct}
{\mcitedefaultendpunct}{\mcitedefaultseppunct}\relax
\EndOfBibitem
\bibitem[Day \latin{et~al.}(2010)Day, Paschek, and
  Garcia]{https://doi.org/10.1002/prot.22702}
Day,~R.; Paschek,~D.; Garcia,~A.~E. Microsecond Simulations of the
  Folding/Unfolding Thermodynamics of the Trp-Cage Miniprotein. \emph{Proteins} \textbf{2010}, \emph{78},
  1889--1899\relax
\mciteBstWouldAddEndPuncttrue
\mciteSetBstMidEndSepPunct{\mcitedefaultmidpunct}
{\mcitedefaultendpunct}{\mcitedefaultseppunct}\relax
\EndOfBibitem
\bibitem[Maffucci \latin{et~al.}(2020)Maffucci, Laage, Sterpone, and
  Stirnemann]{https://doi.org/10.1002/chem.202003018}
Maffucci,~I.; Laage,~D.; Sterpone,~F.; Stirnemann,~G. Cover Feature: Thermal
  Adaptation of Enzymes: Impacts of Conformational Shifts on Catalytic
  Activation Energy and Optimum Temperature. \emph{Chem. Eur. J} \textbf{2020}, \emph{26}, 9657--9657\relax
\mciteBstWouldAddEndPuncttrue
\mciteSetBstMidEndSepPunct{\mcitedefaultmidpunct}
{\mcitedefaultendpunct}{\mcitedefaultseppunct}\relax
\EndOfBibitem
\bibitem[Dabin and Stirnemann(2023)Dabin, and Stirnemann]{dabin_toward_2023}
Dabin,~A.; Stirnemann,~G. Toward a Molecular Mechanism of Complementary {RNA}
  Duplexes Denaturation. \emph{Journal of Physical Chemistry B} \textbf{2023},
  \emph{127}, 6015--6028\relax
\mciteBstWouldAddEndPuncttrue
\mciteSetBstMidEndSepPunct{\mcitedefaultmidpunct}
{\mcitedefaultendpunct}{\mcitedefaultseppunct}\relax
\EndOfBibitem
\bibitem[Dabin and Stirnemann(2024)Dabin, and Stirnemann]{dabin_atomistic_2024}
Dabin,~A.; Stirnemann,~G. Atomistic Simulations of {RNA} Duplex Thermal
  Denaturation: Sequence- and Forcefield-Dependence. \emph{Biophys.
  Chem.} \textbf{2024}, \emph{307}, 107167\relax
\mciteBstWouldAddEndPuncttrue
\mciteSetBstMidEndSepPunct{\mcitedefaultmidpunct}
{\mcitedefaultendpunct}{\mcitedefaultseppunct}\relax
\EndOfBibitem
\bibitem[Forget \latin{et~al.}(2024)Forget, Juill{\'e}, Dubou{\'e}-Dijon, and
  Stirnemann]{forget2024simulation}
Forget,~S.; Juill{\'e},~M.; Dubou{\'e}-Dijon,~E.; Stirnemann,~G.
  Simulation-Guided Conformational Space Exploration to Assess Reactive
  Conformations of a Ribozyme. \emph{J. Chem. Theo. Comput.} \textbf{2024}, \emph{20}, 6263--6277\relax
\mciteBstWouldAddEndPuncttrue
\mciteSetBstMidEndSepPunct{\mcitedefaultmidpunct}
{\mcitedefaultendpunct}{\mcitedefaultseppunct}\relax
\EndOfBibitem
\bibitem[Zinovjev \latin{et~al.}(2024)Zinovjev, Gu{\'e}non, Ramos-Guzm{\'a}n,
  Ruiz-Pern{\'\i}a, Laage, and Tu{\~n}{\'o}n]{zinovjev2024activation}
Zinovjev,~K.; Gu{\'e}non,~P.; Ramos-Guzm{\'a}n,~C.~A.; Ruiz-Pern{\'\i}a,~J.~J.;
  Laage,~D.; Tu{\~n}{\'o}n,~I. Activation and Friction in Enzymatic Loop
  Opening and Closing Dynamics. \emph{Nat. Commun} \textbf{2024},
  \emph{15}, 2490\relax
\mciteBstWouldAddEndPuncttrue
\mciteSetBstMidEndSepPunct{\mcitedefaultmidpunct}
{\mcitedefaultendpunct}{\mcitedefaultseppunct}\relax
\EndOfBibitem
\bibitem[Sohl-Dickstein \latin{et~al.}(2015)Sohl-Dickstein, Weiss,
  Maheswaranathan, and Ganguli]{sohl2015deep}
Sohl-Dickstein,~J.; Weiss,~E.; Maheswaranathan,~N.; Ganguli,~S. Deep
  Unsupervised Learning Using Nonequilibrium Thermodynamics.
  \emph{ICML} \textbf{2015},
  2256--2265\relax
\mciteBstWouldAddEndPuncttrue
\mciteSetBstMidEndSepPunct{\mcitedefaultmidpunct}
{\mcitedefaultendpunct}{\mcitedefaultseppunct}\relax
\EndOfBibitem
\bibitem[Lindahl()]{gromacs20194}
Lindahl, GROMACS User Manual version 2019.4, http://www.gromacs.org. \relax
\mciteBstWouldAddEndPunctfalse
\mciteSetBstMidEndSepPunct{\mcitedefaultmidpunct}
{}{\mcitedefaultseppunct}\relax
\EndOfBibitem
\bibitem[Honda \latin{et~al.}(2008)Honda, Akiba, Kato, Sawada, Sekijima,
  Ishimura, Ooishi, Watanabe, Odahara, and Harata]{honda2008crystal}
Honda,~S.; Akiba,~T.; Kato,~Y.~S.; Sawada,~Y.; Sekijima,~M.; Ishimura,~M.;
  Ooishi,~A.; Watanabe,~H.; Odahara,~T.; Harata,~K. Crystal Structure of a
  Ten-Amino Acid Protein. \emph{J. Am. Chem. Soc}
  \textbf{2008}, \emph{130}, 15327--15331\relax
\mciteBstWouldAddEndPuncttrue
\mciteSetBstMidEndSepPunct{\mcitedefaultmidpunct}
{\mcitedefaultendpunct}{\mcitedefaultseppunct}\relax
\EndOfBibitem
\bibitem[Lindorff-Larsen \latin{et~al.}(2010)Lindorff-Larsen, Piana, Palmo,
  Maragakis, Klepeis, Dror, and Shaw]{lindorff2010improved}
Lindorff-Larsen,~K.; Piana,~S.; Palmo,~K.; Maragakis,~P.; Klepeis,~J.~L.;
  Dror,~R.~O.; Shaw,~D.~E. Improved Side-Chain Torsion Potentials for the Amber
  ff99SB Protein Force Field. \emph{Proteins} \textbf{2010}, \emph{78}, 1950--1958\relax
\mciteBstWouldAddEndPuncttrue
\mciteSetBstMidEndSepPunct{\mcitedefaultmidpunct}
{\mcitedefaultendpunct}{\mcitedefaultseppunct}\relax
\EndOfBibitem
\bibitem[Parrinello and Rahman(1981)Parrinello, and
  Rahman]{parrinello1981polymorphic}
Parrinello,~M.; Rahman,~A. Polymorphic Transitions in Single Crystals: a New
  Molecular Dynamics Method. \emph{J. Appl. Phys} \textbf{1981},
  \emph{52}, 7182--7190\relax
\mciteBstWouldAddEndPuncttrue
\mciteSetBstMidEndSepPunct{\mcitedefaultmidpunct}
{\mcitedefaultendpunct}{\mcitedefaultseppunct}\relax
\EndOfBibitem
\bibitem[Nos{\'e} and Klein(1983)Nos{\'e}, and Klein]{nose1983constant}
Nos{\'e},~S.; Klein,~M. Constant Pressure Molecular Dynamics for Molecular
  Systems. \emph{Mol. Phys.} \textbf{1983}, \emph{50}, 1055--1076\relax
\mciteBstWouldAddEndPuncttrue
\mciteSetBstMidEndSepPunct{\mcitedefaultmidpunct}
{\mcitedefaultendpunct}{\mcitedefaultseppunct}\relax
\EndOfBibitem
\bibitem[Bussi \latin{et~al.}(2007)Bussi, Donadio, and
  Parrinello]{bussi2007canonical}
Bussi,~G.; Donadio,~D.; Parrinello,~M. Canonical Sampling through Velocity
  Rescaling. \emph{J. Chem. Phys.} \textbf{2007}, \emph{126},
  014101\relax
\mciteBstWouldAddEndPuncttrue
\mciteSetBstMidEndSepPunct{\mcitedefaultmidpunct}
{\mcitedefaultendpunct}{\mcitedefaultseppunct}\relax
\EndOfBibitem
\bibitem[Tribello \latin{et~al.}(2025)Tribello, Bonomi, Bussi, Camilloni,
  Armstrong, Arsiccio, Aureli, Ballabio, Bernetti, Bonati, Brookes, Brotzakis,
  Capelli, Ceriotti, Chan, Cossio, Dasetty, Donadio, Ensing, Ferguson, Fraux,
  Gale, Gervasio, Giorgino, Herringer, Hocky, Hoff, Invernizzi,
  Languin-Catto{\"e}n, Leone, Limongelli, Lopez-Acevedo, Marinelli,
  Febrer~Martinez, Masetti, Mehdi, Michaelides, Murtada, Parrinello, Piaggi,
  Pietropaolo, Pietrucci, Pipolo, Pritchard, Raiteri, Raniolo, Rapetti, Rizzi,
  Rydzewski, Salvalaglio, Schran, Seal, Shayesteh~Zadeh, Silva, Spiwok,
  Stirnemann, Sucerquia, Tiwary, Valsson, Vendruscolo, Voth, White, and
  Wu]{Tribello:2025aa}
Tribello,~G.~A. \latin{et~al.}  PLUMED Tutorials: A Collaborative,
  Community-Driven Learning Ecosystem. \emph{J Chem Phys} \textbf{2025},
  \emph{162}\relax
\mciteBstWouldAddEndPuncttrue
\mciteSetBstMidEndSepPunct{\mcitedefaultmidpunct}
{\mcitedefaultendpunct}{\mcitedefaultseppunct}\relax
\EndOfBibitem
\bibitem[Lindorff-Larsen \latin{et~al.}(2011)Lindorff-Larsen, Piana, Dror, and
  Shaw]{lindorff2011fast}
Lindorff-Larsen,~K.; Piana,~S.; Dror,~R.~O.; Shaw,~D.~E. How Fast-Folding
  Proteins Fold. \emph{Science} \textbf{2011}, \emph{334}, 517--520\relax
\mciteBstWouldAddEndPuncttrue
\mciteSetBstMidEndSepPunct{\mcitedefaultmidpunct}
{\mcitedefaultendpunct}{\mcitedefaultseppunct}\relax
\EndOfBibitem
\bibitem[plu(2019)]{plumed2019promoting}
The PLUMED Consortium, Promoting Transparency and Reproducibility in Enhanced
  Molecular Simulations. \emph{Nat. Methods} \textbf{2019}, \emph{16},
  670--673\relax
\mciteBstWouldAddEndPuncttrue
\mciteSetBstMidEndSepPunct{\mcitedefaultmidpunct}
{\mcitedefaultendpunct}{\mcitedefaultseppunct}\relax
\EndOfBibitem
\bibitem[Tribello \latin{et~al.}(2014)Tribello, Bonomi, Branduardi, Camilloni,
  and Bussi]{tribello2014plumed}
Tribello,~G.~A.; Bonomi,~M.; Branduardi,~D.; Camilloni,~C.; Bussi,~G. PLUMED 2:
  New Feathers for an Old Bird. \emph{Comput. Phys. Commun}
  \textbf{2014}, \emph{185}, 604--613\relax
\mciteBstWouldAddEndPuncttrue
\mciteSetBstMidEndSepPunct{\mcitedefaultmidpunct}
{\mcitedefaultendpunct}{\mcitedefaultseppunct}\relax
\EndOfBibitem
\bibitem[Kumar \latin{et~al.}(1992)Kumar, Rosenberg, Bouzida, Swendsen, and
  Kollman]{kumar1992weighted}
Kumar,~S.; Rosenberg,~J.~M.; Bouzida,~D.; Swendsen,~R.~H.; Kollman,~P.~A. The
  Weighted Histogram Analysis Method for Free-Energy Calculations on
  Biomolecules. I. The Method. \emph{J. Comput. Chem}
  \textbf{1992}, \emph{13}, 1011--1021\relax
\mciteBstWouldAddEndPuncttrue
\mciteSetBstMidEndSepPunct{\mcitedefaultmidpunct}
{\mcitedefaultendpunct}{\mcitedefaultseppunct}\relax
\EndOfBibitem
\bibitem[Grossfield(2023)]{grossfield2023wham}
Grossfield,~A. WHAM: Weighted Histogram Analysis Method for Analyzing Umbrella
  Sampling Simulation Data. \textbf{2023}, \relax
\mciteBstWouldAddEndPunctfalse
\mciteSetBstMidEndSepPunct{\mcitedefaultmidpunct}
{}{\mcitedefaultseppunct}\relax
\EndOfBibitem
\bibitem[Paul~Bauer(2022)]{gromacs20224}
Paul~Bauer,~E.~L.,~Berk~Hess GROMACS 2022.4 Source code (2022.4). Zenodo.
  https://doi.org/10.5281/zenodo.7323393. \textbf{2022}, \relax
\mciteBstWouldAddEndPunctfalse
\mciteSetBstMidEndSepPunct{\mcitedefaultmidpunct}
{}{\mcitedefaultseppunct}\relax
\EndOfBibitem
\bibitem[Zhang(2001)]{zhang2001protein}
Zhang,~Z.-Y. Protein Tyrosine Phosphatases: Prospects for Therapeutics.
  \emph{Curr Opin Chem Biol.} \textbf{2001}, \emph{5},
  416--423\relax
\mciteBstWouldAddEndPuncttrue
\mciteSetBstMidEndSepPunct{\mcitedefaultmidpunct}
{\mcitedefaultendpunct}{\mcitedefaultseppunct}\relax
\EndOfBibitem
\bibitem[Feldhammer \latin{et~al.}(2013)Feldhammer, Uetani, Miranda-Saavedra,
  and Tremblay]{feldhammer2013ptp1b}
Feldhammer,~M.; Uetani,~N.; Miranda-Saavedra,~D.; Tremblay,~M.~L. PTP1B: a
  Simple Enzyme for a Complex World. \emph{Crit. Rev. Biochem. Mol. Biol} \textbf{2013}, \emph{48}, 430--445\relax
\mciteBstWouldAddEndPuncttrue
\mciteSetBstMidEndSepPunct{\mcitedefaultmidpunct}
{\mcitedefaultendpunct}{\mcitedefaultseppunct}\relax
\EndOfBibitem
\bibitem[Whittier \latin{et~al.}(2013)Whittier, Hengge, and
  Loria]{whittier2013conformational}
Whittier,~S.~K.; Hengge,~A.~C.; Loria,~J.~P. Conformational Motions Regulate
  Phosphoryl Transfer in Related Protein Tyrosine Phosphatases. \emph{Science}
  \textbf{2013}, \emph{341}, 899--903\relax
\mciteBstWouldAddEndPuncttrue
\mciteSetBstMidEndSepPunct{\mcitedefaultmidpunct}
{\mcitedefaultendpunct}{\mcitedefaultseppunct}\relax
\EndOfBibitem
\bibitem[Shen \latin{et~al.}(2021)Shen, Crean, Johnson, Kamerlin, and
  Hengge]{shen2021single}
Shen,~R.; Crean,~R.~M.; Johnson,~S.~J.; Kamerlin,~S.~C.; Hengge,~A.~C. Single
  Residue on the WPD-Loop Affects the pH Dependency of Catalysis in Protein
  Tyrosine Phosphatases. \emph{JACS Au} \textbf{2021}, \emph{1}, 646--659\relax
\mciteBstWouldAddEndPuncttrue
\mciteSetBstMidEndSepPunct{\mcitedefaultmidpunct}
{\mcitedefaultendpunct}{\mcitedefaultseppunct}\relax
\EndOfBibitem
\bibitem[Zinovjev and Tunon(2017)Zinovjev, and Tunon]{zinovjev2017adaptive}
Zinovjev,~K.; Tunon,~I. Adaptive Finite Temperature String Method in Collective
  Variables. \emph{J. Phys. Chem. A} \textbf{2017},
  \emph{121}, 9764--9772\relax
\mciteBstWouldAddEndPuncttrue
\mciteSetBstMidEndSepPunct{\mcitedefaultmidpunct}
{\mcitedefaultendpunct}{\mcitedefaultseppunct}\relax
\EndOfBibitem
\bibitem[Davis \latin{et~al.}(2011)Davis, Lii, and Politis]{davis2011remarks}
Davis,~R.~A.; Lii,~K.-S.; Politis,~D.~N. Remarks on Some Nonparametric
  Estimates of a Density Function. \emph{Selected Works of Murray Rosenblatt}
  \textbf{2011}, 95--100\relax
\mciteBstWouldAddEndPuncttrue
\mciteSetBstMidEndSepPunct{\mcitedefaultmidpunct}
{\mcitedefaultendpunct}{\mcitedefaultseppunct}\relax
\EndOfBibitem
\bibitem[Parzen(1962)]{parzen1962estimation}
Parzen,~E. On Estimation of a Probability Density Function and Mode. \emph{Ann. Math. Stat} \textbf{1962}, \emph{33}, 1065--1076\relax
\mciteBstWouldAddEndPuncttrue
\mciteSetBstMidEndSepPunct{\mcitedefaultmidpunct}
{\mcitedefaultendpunct}{\mcitedefaultseppunct}\relax
\EndOfBibitem
\bibitem[Lee(2025)]{lee25}
Lee,~S.; Wang, R.; Herron, L.; Tiwary, P. Exponentially Tilted Thermodynamic Maps (expTM): Predicting Phase Transitions Across Temperature, Pressure, and Chemical Potential. \emph{arXiv} \textbf{2025} 2503.12080\relax
\mciteBstWouldAddEndPuncttrue
\mciteSetBstMidEndSepPunct{\mcitedefaultmidpunct}
{\mcitedefaultendpunct}{\mcitedefaultseppunct}\relax
\EndOfBibitem
\end{mcitethebibliography}
\providecommand{\latin}[1]{#1}
\makeatletter
\providecommand{\doi}
  {\begingroup\let\do\@makeother\dospecials
  \catcode`\{=1 \catcode`\}=2 \doi@aux}
\providecommand{\doi@aux}[1]{\endgroup\texttt{#1}}
\makeatother
\providecommand*\mcitethebibliography{\thebibliography}
\csname @ifundefined\endcsname{endmcitethebibliography}
  {\let\endmcitethebibliography\endthebibliography}{}

\newpage 

\begin{figure}[t!]
\centering
\includegraphics[width=2.0\linewidth]{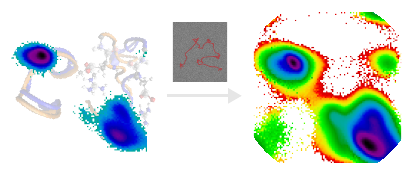} 
Table of content figure
\label{fig:toc}
\end{figure}

\end{document}